\documentclass[a4paper,fleqn,usenatbib]{mnras}

\usepackage{newtxtext,newtxmath}

\usepackage[T1]{fontenc}
\usepackage{ae,aecompl}

\usepackage{graphicx}	
\usepackage{amsmath}	
 \usepackage{amssymb}	
\usepackage{color}

\title[Mass-acceleration degeneracy for the DWDs in the LISA band]{Degeneracy between mass and peculiar acceleration for the double white dwarfs in the LISA band }

\author[Z. Xuan, P. Peng and X. Chen]{Zeyuan Xuan$^{1}$, Peng Peng$^{1}$\thanks{E-mail: \href{mail to: peng.p@pku.edu.cn}{peng.p@pku.edu.cn}} and Xian Chen$^{1,2}\thanks{E-mail: \href{mail to: xian.chen@pku.edu.cn}{xian.chen@pku.edu.cn}} $
\\
$^{1}$Astronomy Department, School of Physics, Peking University, 100871 Beijing, China
\\
$^{2}$Kavli Institute for Astronomy and Astrophysics at Peking University, 100871 Beijing, China
}

\date{Accepted XXX. Received YYY; in original form ZZZ}

\pubyear{2020}

\begin{document}
\label{firstpage}
\pagerange{\pageref{firstpage}--\pageref{lastpage}}
\maketitle

\begin{abstract}
Mass and distance are fundamental quantities to measure in gravitational-wave
(GW) astronomy.  However, recent studies suggest that the measurement may be
biased due to the acceleration of GW source. Here we develop an analytical
method to quantify such a bias induced by a tertiary on a double white dwarf
(DWD), since DWDs are the most common GW sources in the milli-Hertz band. We show
that in a large parameter space the mass is degenerate with the peculiar
acceleration, so that from the waveform we can only retrieve a mass of ${\cal M}(1+\Gamma)^{3/5}$,
where ${\cal M}$ is the real chirp mass of the DWD and $\Gamma$ is a dimensionless
factor proportional to the peculiar acceleration.  Based on our analytical
method, we conduct mock observation of DWDs by the Laser Interferometer Space
Antenna (LISA). We find that in about $9\%$ of the cases the measured chirp
mass is biased due to the presence of a tertiary by $(5-30)\%$.
Even more extreme cases are found in about a dozen DWDs and they may be misclassified as double neutron stars,
binary black holes, DWDs undergoing mass transfer, or even binaries containing
lower-mass-gap objects and
primordial black holes.  The bias in mass also affects
the measurement of distance, resulting in a seemingly over-density of DWDs
within a heliocentric distance of $1$ kpc as well as beyond $100$ kpc.  
Our result highlights the necessity of modeling
the astrophysical environments of GW sources to retrieve their correct
physical parameters.  
\end{abstract}




\section{Introduction}\label{sec:intro}

Measuring mass is a fundamental problem in Gravitational Wave (GW) astronomy.
For sources such as binary black holes (BBHs), binary neutron stars (BNSs) and
double white dwarfs (DWDs), the information of mass can be extracted from the
inspiral waveform \citep[see][for a review]{2009LRR....12....2S}.  A
prerequisite is that the orbital dynamics of the binary is dominated by GW
radiation and the GWs propagate freely to the observer.  In this ideal situation,
the chirp signal, i.e., an increase of GW frequency $f$ with time, is uniquely
determined by a quantity called ``chirp mass''. It depends on the masses $m_1$
and $m_2$ of the two compact objects as ${\cal
M}=(m_1m_2)^{3/5}/(m_1+m_2)^{1/5}$ \citep{1994PhRvD..49.2658C}. As a result of
such a dependence, ${\cal M}$ can be derived from $f$ and its time derivative
$\dot{f}$ (``chirp rate'' hereafter).

In reality, however, the astrophysical environment of the source could affect
the propagation of GWs or perturb the binary orbit. As a result, the chirp
signal is distorted and the measurement of the mass could be biased
\citep[see][for a summary]{chen20envi}. For example, it is well known that the
expansion of the universe could stretch the waveform and make the chirp mass
appear bigger \citep{2009LRR....12....2S,broadhurst18,smith18}.  Such an effect
is called the ``mass-redshift degeneracy''. It has been generalized recently to
include the Doppler and gravitational redshift, to show that the GW sources in
the vicinity of supermassive black holes (SMBHs) would also appear more massive
\citep{chen19}. Moreover, it has been pointed out that the gas around BBHs, by
accelerating the orbital shrinkage and increasing $\dot{f}$, could also lead to
an overestimation of the chirp masses \citep{chen19gas,caputo20,chen20gas}.

Peculiar acceleration is another environmental factor which could bias the
measurement of the mass of a GW source \citep{2017PhRvD..95d4029B}.  It is
caused by a variation of the centre-of-mass (CoM) velocity of the source. Such
a variation could be induced by the gravitational potential of the environment,
such as a star cluster,  
and more likely by a tertiary object since many
binaries are in triple systems.  The consequence is a time-dependent Doppler
shift of the GW frequency (\citealt{2017PhRvD..95d4029B}, also see
\citealt{2008ApJ...677L..55S} for additional effects when the orbital period of
the triple is particularly short).  Earlier studies focused on using the
frequency shift to  identify the GW sources in triple systems, such as an
extreme-mass-ratio or intermediate-mass-ratio inspiral with a SMBH tertiary
\citep{2011PhRvD..83d4030Y,Deme20}, a BBH orbiting a star or a SMBH
\citep{2017PhRvD..96f3014I,2017ApJ...834..200M,2019ApJ...878...75R,2019MNRAS.488.5665W,2020PhRvD.101f3002T},
and a DWD accompanied by a star or a planet
\citep{2008ApJ...677L..55S,2018PhRvD..98f4012R,2018arXiv181203438S,
2019NatAs...3..858T}. However, more recent works revealed that the effect due
to peculiar acceleration may be indistinguishable from a normal chirp signal
when the orbital period of the tertiary is much longer than the observational
period. In this case the additional frequency shift induced by the peculiar
acceleration could mislead our measurement of $\dot{f}$ and, in turn,  cause an
bias in the measurement of the chirp mass
\citep{2018PhRvD..98f4012R,2020PhRvD.101f3002T}. 

Such a bias should be more prominent for DWDs than for BBHs or DNSs.  First,
DWDs have smaller chirp masses, and hence a smaller chirp rate $\dot{f}$ since
it depends on ${\cal M}^{5/3}$. 
Consequently, they are more easily affected by the frequency shift
induced by
peculiar acceleration \citep{chen20envi}.  Second, DWDs are the most numerous
sources in the milli-Hertz (mHz) GW band. They are among the prime targets of
the
Laser Interferometer
Space Antenna
\citep[LISA,][]{nelemans01,2002MNRAS.333..469S,2017arXiv170200786A}, while BBHs
and DNSs are more easily detected by high-frequency detectors, such as LIGO and
Virgo \citep{abbott19population}. The relatively low frequency of DWDs further reduces the
chirp rate since $\dot{f}\propto f^{11/3}$. Third, both theoretical
models \citep{2016ComAC...3....6T} and observations \citep{2017A&A...602A..16T,
2019MNRAS.483..901P,lagos20} suggest that a large fraction of DWDs are in
triples.  Moreover, observations also show that in the Milky Way more than half
of the solar-type binaries which have a period shorter than $3$ days are in
triple systems \citep{2006AJ....131.2986P,2008msah.conf..129T}, and a fraction
of $\sim 30$ per cent of the massive stars around the
SMBH in the Galactic Centre are in binaries \citep{2014ApJ...782..101P}. After these
binary stars evolve into DWDs, it is likely that they are surrounded by tertiaries
\citep{2013MNRAS.430.2262H, 2019ApJ...878...58S}.

Indeed, \citet{2018PhRvD..98f4012R} found cases in which 
a DWD in a triple system is misidentified by LISA as a 
isolated DWD with a slight different mass.  They also pointed out that the error in
the chirp mass would propagate into the estimation of the
luminosity distance.  However, it is still unclear how common
such a bias is among the DWD population in the Milky Way.  Given the potential usage
of the LISA DWDs in mapping the Milky-Way structure
\citep{cooray05,adams12,2019MNRAS.483.5518K,breivik20}, we will study in this paper the
impact of peculiar acceleration on the measurement of their chirp masses and
distances. 

The paper is organized as follows. In Section~\ref{sec:theory}, we derive
analytical formulae for the biases in mass and distance induced by the peculiar
acceleration.  In Section~\ref{sec:confusion}, we explore the parameter space
of DWDs in which a tertiary induces a significant bias but remains undetectable
by LISA.  We verify the above analytical results using the numerical
matched-filtering technique in Section~\ref{sec:numerical}. Then in
Section~\ref{sec:population}, we use a bifurcation model to general a population of
DWDs in the LISA band with a realistic fraction of triples.  We conduct mock
LISA observation of the simulated DWD population in
Section~\ref{sec:observation}. Finally, we summarize our results and conclude
in Section~\ref{sec:conclusion}.

\section{The effect of peculiar acceleration}\label{sec:theory}

The general effect of peculiar acceleration has been discussed in
\citet{2017PhRvD..95d4029B}. Here we focus on the effect in triple systems
containing DWDs.  In particular, we derive analytical formulae for the
resulting biases in the measurement of the chirp mass and distance. These
formulae are useful in our later analysis of a large sample of DWDs. 

LISA estimates the chirp mass ${\cal M}$ of a DWD based on two observables, the
frequency $f_e$ and its time derivative $\dot{f}_e$ (``chirp rate'' hereafter), where
the subscript $e$ denotes the quantities in the rest frame of the source.
The details of the method can be found in \citet{1994PhRvD..49.2658C} and the
result is
\begin{equation}
	{\cal M}:=\frac{(m_1m_2)^{3/5}}{(m_1+m_2)^{1/5}}
	={ \left(
\frac{5 c^5}{96 \pi^{8/3} G^{5/3} } \right) }^{3/5} f_e^{-11/5} \dot{f}_e^{3/5},
\label{eq:chirpmass-f}
\end{equation}
which assumes a circular orbit for the binary. This result suggests that
the chirp signal evolves on a timescale of
\begin{equation}
	\tau_{\rm gw}:=\frac{f_e}{\dot{f}_e}\simeq2.2\times10^6\,
	\left(\frac{{\cal M}}{0.3\,M_\odot}\right)^{-5/3}
	\left(\frac{f_e}{3\,{\rm mHz}}\right)^{-8/3}\,{\rm years}.
\end{equation}
Note that this timescale is much longer than the mission duration of LISA, which
is about $4-5$ years \citep{2017arXiv170200786A}.

Using ${\cal M}$ and the GW amplitude $h$, which is the third observable, one can further infer
the distance of the DWD as 
\begin{equation}
d=\frac{4G}{c^2}\frac{{\cal M}}{h}\left(\frac{G}{c^3}\pi f_e {\cal M}\right)^{2/3}.\label{eq:d}
\end{equation}
In the last equation we have omitted the uncertainty induced by the inclination
of the binary because in principle it can be eliminated by measuring the
relative strength of the two GW polarizations.

Not all DWDs are suitable for LISA to measure their masses.
The chirp rate should exceed a threshold to induce a detectable frequency shift. 
\citet{2002MNRAS.333..469S} showed that the threshold is
$\Delta \dot{f} = 6\sqrt{5}/(\pi  \rho \tau_{\rm obs}^{2})$, 
where $\rho$ denotes the signal-to-noise ratio (SNR), and
$\tau_{\rm obs}$ is the duration of the observation.
Such a criterion is equivalent to the requirement 
\begin{equation}
	f_e\ga1.67\,{\rm mHz}\,\left(\frac{{\cal M}}{0.3\,M_\odot}\right)^{-5/11}
	\left(\frac{\rho}{100}\right)^{-3/11}
	\left(\frac{\tau_{\rm obs}}{4\,{\rm years}}\right)^{-6/11},\label{eq:resolve}
\end{equation}
where a canonical observational period of $4$ years is assumed.  We also
assumed a chirp mass of $\mathcal{M}=0.3\, M_{\sun}$ because it is typical for
He-He and He-CO DWDs, the most common DWDs in the LISA band
\citep{2019MNRAS.490.5888L}.  We have imposed a stringent criterion $\rho=10^2$
for detecting the chirp signal.  Such a SNR corresponds to a distance of  $\sim
2\rm kpc$ to our typical DWD.   Nevertheless, the result is not sensitive to
our choice of $\rho$.   

The variation of the chirp rate, $\ddot{f}_e$, is usually more difficult to
detect due to the long $\tau_{\rm gw}$. For example, when GW radiation
predominates, we have $\ddot{f}_e=11\dot{f}_e/(3\tau_{\rm gw})$. During the
observational period $\tau_{\rm obs}$, such a $\ddot{f}_e$ leads to a small
increase of $\dot{f}_e$ by an amount of $\delta{\dot{f}}\sim
\ddot{f}_e\tau_{\rm obs}$, which is of the order of $\dot{f}_e(\tau_{\rm
obs}/\tau_{\rm gw})$. It is much smaller than $\dot{f}_e$, and hence more
difficult to detect. The implication, which is important for
the later understanding the effect of peculiar acceleration, is that the chirp signal
($f_e$ as a function of $t$) has a negligible curvature, 
i.e., it is almost a straight line in the time-frequency diagram
when the DWD is in the LISA band.
 
Peculiar acceleration affects both the frequency and the chirp rate.  The
frequency is Doppler shifted by a factor of
\begin{equation} 1+z_{\rm dop}=\frac{1+\beta \cos{\theta}}{\sqrt{1-\beta^{2}}},
\end{equation}
where $\beta:=v/c$ denotes the ratio between the CoM velocity of the DWD ($v$)
and the speed of light ($c$), and $\theta$ is the angle between the velocity
vector $\mathbf{v}$ and the line-of-sight $\mathbf{n}$. Note that
both $\mathbf{v}$ and $\mathbf{n}$ are measured in the observer's frame.
Therefore, to the observer the frequency is
\begin{equation} f_o = f_e(1+z_{\rm dop})^{-1}.  \label{eq:doppler}
\end{equation}
The chirp rate is affected even more because of the effect of time dilation, so
that the observed chirp rate is
\begin{align} 
	\dot{f}_o&=\frac{df_o}{dt_o}=\frac{dt}{dt_o}\frac{df_o}{dt}\\
&=\frac{\dot{f}_e}{(1+z_{\rm
dop})^2}+f_o\frac{d}{dt}\left(\frac{\sqrt{1-\beta^2}}{1+\beta\cos\theta}\right).\label{eq:fodotacc}
\end{align}
In the last equation, the first term refers to the effect of a constant
velocity and the second term, which we denote as $\dot{f}_{\rm acc}$, is caused
by peculiar acceleration. Note that $\theta$ is a period function of time as the
DWD orbits around the tertiary.

The apparent chirp mass, which is the only mass scale derivable from the
observed waveform, should be computed from $f_o$ and $\dot{f}_o$ as 
\begin{equation}
	{\cal M}_o
	={ \left(
\frac{5 c^5}{96 \pi^{8/3} G^{5/3} } \right) }^{3/5} f_o^{-11/5} \dot{f}_o^{3/5}.
\label{eq:chirpmass-o}
\end{equation}
To facilitate the computation, we define $\Gamma$ to be the ratio between the
second and the first term in Equation~(\ref{eq:fodotacc}), so that
\begin{equation}
\dot{f}_o=\dot{f}_e(1+z_{\rm dop})^{-2}(1+\Gamma).
\end{equation}
Note that $\Gamma$ could be either positive or negative. It follows that
\begin{equation}
	{\cal M}_o={\cal M}(1+z_{\rm dop})(1+\Gamma)^{3/5}.\label{eq:Mo}
\end{equation}
The last equation suggests that the mass estimated from the chirp signal
is biased. The bias is caused not only by 
the Doppler effect but also the peculiar acceleration. 

As for the distance,
we note that the observer could only use $f_o$ and ${\cal M}_o$
in Equation~(\ref{eq:d}). As a result, the distance appears to be
\begin{equation}
	d_o=d(1+z_{\rm dop})(1+\Gamma).\label{eq:do}
\end{equation}
It is clear that the apparent distance of the source is also biased
in the presence of a peculiar acceleration.

We now show that the value of $|\Gamma|$ could be much greater than 
that of $|z_{\rm dop}|$.
Before doing the calculation, we
should first specify the parameters of the tertiary.  For simplicity, we assume
a circular orbit for the tertiary. In this case, the $\beta$ parameter can be calculated
with
\begin{align}
\beta&=\frac{m_3}{m_{12}+m_3}\sqrt{\frac{G(m_{12}+m_3)}{r\,c^2}},\\
&\simeq2.6\times10^{-5}\left(\frac{m_3}{m_{12}+m_3}\right)
\left(\frac{m_{12}+m_3}{2\,M_\odot}\right)^{1/2}
\left(\frac{r}{30\,{\rm AU}}\right)^{-1/2},\\
&\simeq1.4\times10^{-3}\left(\frac{m_3}{m_{12}+m_3}\right)
\left(\frac{m_{12}+m_3}{4\times10^{6}\,M_\odot}\right)^{1/2}
\left(\frac{r}{0.1\,{\rm pc}}\right)^{-1/2}.
\end{align}
where $m_{12}=m_1+m_2$, $m_3$ is the mass of the tertiary and $r$ is the
distance from the tertiary to the CoM of the DWD.  We find that the value of
$\beta$ is small for the triple systems of our interest.  For example, if we
use $m_{12}=1\,M_\odot$, $m_3=1\,M_\odot$, and $r=30$ AU to represent the DWDs
around main-sequence stars, we find that $\beta\simeq1.3\times10^{-5}$. 
If we
increase $m_3$ to $4\times10^6\,M_\odot$ and $r$ to $0.1$ pc, to represent the
DWDs around the SMBH in the Milky Way \citep{2009ApJ...692.1075G}, we find that
$\beta\simeq1.4\times10^{-3}$.  Therefore, we conclude that $|z_{\rm dop}|$ is
of the order of $10^{-5}-10^{-3}$ for our problem.  To estimate $\Gamma$, we first notice
that the second term on the right-hand-side of Equation~(\ref{eq:fodotacc}) is
of the order of $\dot{f}_{\rm acc}\sim Gm_3f_o\cos i/(r^2c)$, where $i\in[0,\pi]$ is the inclination
angle between the line-of-sight and the orbital plane of the DWD.  
Using such an approximation, we find that
\begin{align}
        |\Gamma|&\simeq1.5\left(\frac{f_e}{3{\rm mHz}}\right)^{-8/3}
        \left(\frac{{\cal M}}{0.3M_\odot}\right)^{-5/3}
        \left(\frac{m_3}{M_\odot}\right)
        \left(\frac{r}{30{\rm AU}}\right)^{-2},\label{eq:Gamma1}\\
&\simeq14\left(\frac{f_e}{3{\rm mHz}}\right)^{-8/3}
        \left(\frac{{\cal M}}{0.3M_\odot}\right)^{-5/3}
        \left(\frac{m_3}{4\times10^{6}M_\odot}\right)
        \left(\frac{r}{0.1{\rm pc}}\right)^{-2}.\label{eq:Gamma2}
\end{align}
In the above two equations, we have assumed $\cos i=1$ for
simplicity, but later in this paper we will consider a random
distribution of $i$.  

\begin{figure*}
\begin{minipage}[t]{0.45\linewidth}
\centering
\includegraphics[width=3in]{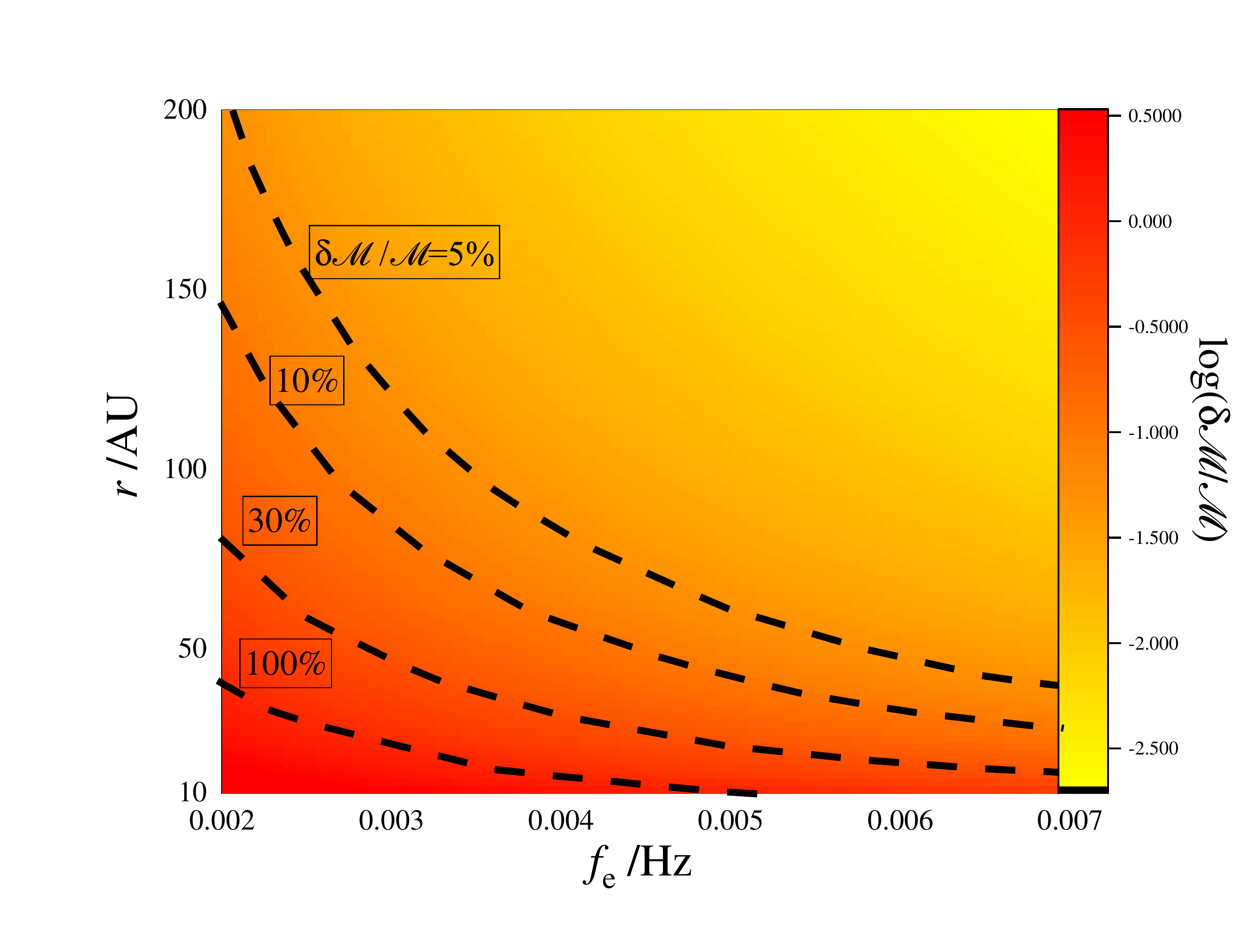}
\end{minipage}
\begin{minipage}[t]{0.45\linewidth}
\centering
\includegraphics[width=3in]{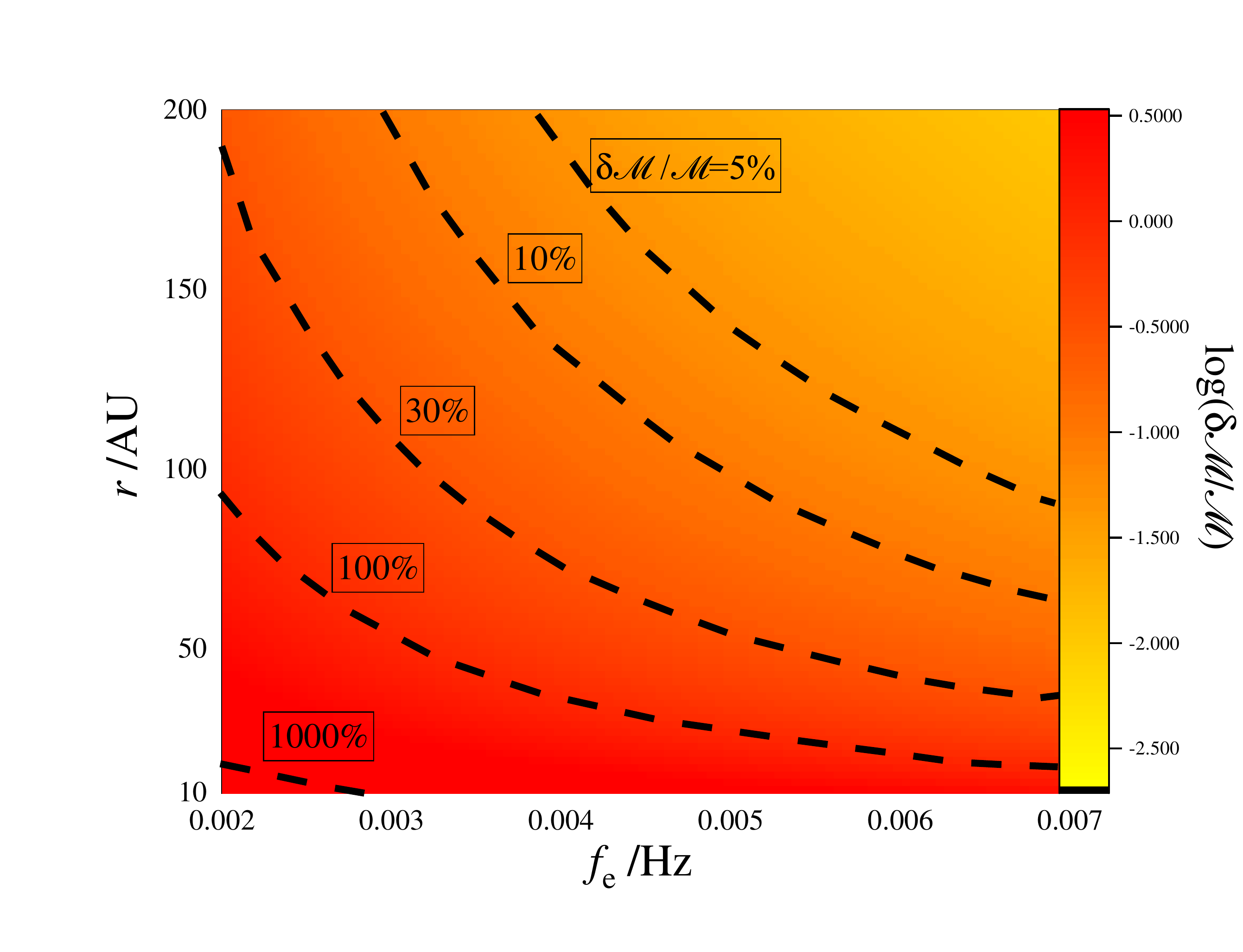}
\end{minipage}
\caption{Bias of the chirp mass induced by the peculiar acceleration. The result is shown
as a function of the GW frequency $f_e$ and the distance $r$ between the DWD and the 
tertiary star. The left panel corresponds to a stellar mass of $m_3=1\,M_\odot$ and the right
one corresponds to $m_3=5\,M_\odot$. The dashed lines show the contours of 
	different values of $\delta {\cal M}/{\cal M}$. In both cases, the chirp mass of the DWD
is set to ${\cal M}=0.3\,M_\odot$. The outer orbit is circular and aligned with the
line-of-sight.
}
\label{fig:analyticresult}
\end{figure*}

Now it is clear that $|\Gamma|\gg|z_{\rm dop}|$ for the DWDs of our interest.
This result, together with Equation~(\ref{eq:Mo}), imply that peculiar
acceleration predominates the  bias in the measurement of the chirp mass and distance.  
To see more clearly the magnitude of such a bias, we show in
Figure~\ref{fig:analyticresult} the value of $\delta {\cal M}={\cal M}_o-{\cal
M}$. In the plot, we have restricted our calculation to the case ${\cal
M}_o>{\cal M}$, i.e., we have assumed $\Gamma>0$. We adopted the maximum
peculiar acceleration along the orbit.  The result could be regarded as the
upper limit of the bias. In general, we find that the bias is greater when the
GW frequency is lower, the distance from the DWD to the tertiary is smaller,
and the tertiary mass is higher. When $m_3=1\,M_\odot$, the bias could be as
large as ${\cal M}$, and when $m_3=5\,M_\odot$, it could even reach $10\,{\cal
M}$ for small $r$. The latter result implies that we might confuse DWDs with DNSs or BBHs.
In fact, these confusing DWDs may fall in the ``lower mass gap'' between $3$
and $5\,M_\odot$, which in the past was considered to be devoid of compact
objects \citep{2010ApJ...725.1918O, 2011ApJ...741..103F}.
 
\begin{figure}
\centering
\includegraphics[width=3in]{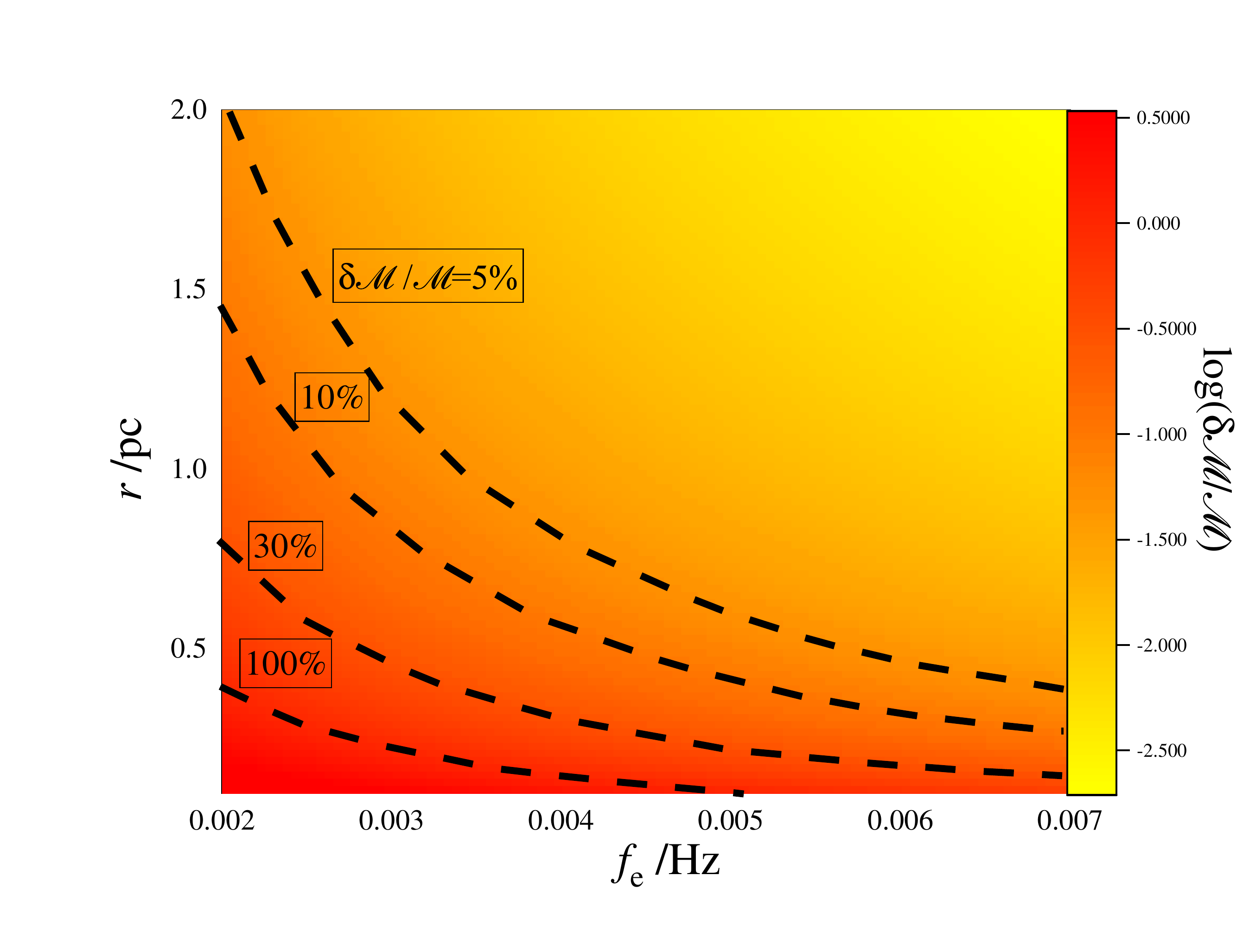}
\caption{The same as Figure \ref{fig:analyticresult}, but the tertiary is a SMBH
	with a mass of $4\times10^6\,M_\odot$.}
\label{fig:SMBH}
\end{figure}

For the DWDs in the vicinity of the SMBH in the Galactic Center, the resulting bias is shown
in Figure~\ref{fig:SMBH}. We find a large parameter space in which the value of 
$\delta{\cal M}/{\cal M}$ exceeds $5\%$. In the lower-left corner of the diagram, 
the bias exceeds $100\%$ of ${\cal M}$. This is the region where the frequency shift
due to peculiar acceleration greatly exceeds what is caused by GW radiation.
Although not shown here, at a distance significantly smaller than $0.1$ pc,
the bias could also exceed $10{\cal M}$.

We note that the DWDs presented in Figure~\ref{fig:SMBH} are dynamically stable.
First, they are stable against the disruptive tidal force of the SMBH, since 
tidal disruption requires that
\begin{equation}
r \la 0.1\, {\rm{AU}}\, \left(\frac{m_3}{4 \times 10^6 \, M_{\odot}}\right)^{1/3} 
	\left(\frac{f_e}{3 \, \rm{mHz}}\right)^{-2/3},
\end{equation}
where $4\times10^{6} \,M_{\sun}$ is the mass of the SMBH in the Galactic Centre
\citep{2009ApJ...692.1075G}.  Second, the secular interaction with the SMBH
also induces a Lidov-Kozai evolution of the DWDs.  However, the timescale is of
the order of
\begin{equation}
T_{\rm{LK}} \sim 2\times 10^{10} {\rm{yr}} 
	\left(\frac{m_3}{4 \times 10^6 M_{\odot}}\right)^{-1} 
	\left(\frac{f_e}{3 \rm{mHz}}\right) 
	\left(\frac{r}{0.1 \rm{pc}}\right)^{3} (1-e_2^2)^{3/2},\label{eq:TKL}
\end{equation}
where $e_2$ is the eccentricity of the orbit of the DWD around the SMBH
\citep[the outer orbit, see][for a review]{2016ARA&A..54..441N} and the mean value is about $0.67$ if the orbits of
the DWDs are thermalized. Therefore, we can neglect the Lidov-Kozai effect
during the observational period of four years of LISA. 
Equation~(\ref{eq:TKL}) also indicates that the DWDs inside a radius of $0.1$ pc
from the SMBH would have been depleted because the Lidov-Kozai effect could have
driven them to merge. For this reason, we do not consider the DWDs with $r<0.1$ pc. 
Third, the specific
internal kinetic energies of the DWDs are higher than the specific kinetic energies
of the background stars by a factor of 
\begin{equation}
\frac{Gm_1m_2}{(m_1+m_2)a_1 \sigma_*^2 } = 3 
	\left(\frac{m_1}{0.34 \, M_{\odot}}\right)^{2/3} 
	\left(\frac{f_e}{3 \, \rm{mHz}}\right)^{2/3} 
	\left(\frac{\sigma_*}{200 \, \rm{km}\, \rm{s}^{-1}}\right)^{-2},
\end{equation}
where $\sigma_*$ denotes the velocity dispersion of the background stars.
Such binaries are ``hard'', so that they would not be ionized during the
interactions with the surrounding stars
\citep{1987gady.book.....B}.

\section{Criterion for confusion}\label{sec:confusion}

Now we study whether LISA could distinguish those DWDs affected by peculiar
acceleration. Although an analytical criterion was provided in
\citet{2018PhRvD..98f4012R}, we derive a simplified version here for
completeness, and for our later analysis of the population of DWDs in the Milky
Way. 

First, we note that the effect of peculiar acceleration is important only when
the induced bias, $\delta {\cal M}$, is greater than the measurement error 
of chirp-mass, $\sigma_{\cal M}$.  According to
Equation \ref{eq:chirpmass-f}, the variation of ${\cal M}$ is related to the
variation of $\dot{f}_e$. For this reason,  the condition $|\delta {\cal
M}|<\sigma_{\cal M}$ is equivalent to $|\dot{f}_o-\dot{f}_e| < \Delta \dot{f}$,
where $\Delta \dot{f} = 6\sqrt{5}/\pi \cdot \rho^{-1} \tau_{\rm obs}^{-2}$ is
the aforementioned threshold of detecting a frequency shift by LISA \citep{2002MNRAS.333..469S,
2002ApJ...575.1030T}.  Substituting Equation~(\ref{eq:fodotacc}) for
$\dot{f}_o$, we find that peculiar acceleration is not important when
\begin{equation}
r\ga 150 \, {\rm{AU}}\,  \left( \frac{\rho}{100} \right)^{1/2} 
	\left( \frac{\tau_{\rm obs}}{4 \, {\rm{yr}}} \right) 
	\left( \frac{f_e}{3 \, {\rm{mHz}}} \right)^{1/2}   
	\left( \frac{ m_3 }{1 \, M_{\odot}} \right)^{1/2}, 
\label{eq: upper_limit_r}
\end{equation}
where, again, we have neglected the higher-order terms of $\beta$, 
because of their smallness. For the DWDs in the Galactic Centre,
the criterion becomes
\begin{equation}
r\ga 0.46 \, {\rm{pc}}\,  \left( \frac{\rho}{10} \right)^{1/2} 
        \left( \frac{\tau_{\rm obs}}{4 \, {\rm{yr}}} \right) 
        \left( \frac{f_e}{3 \, {\rm{mHz}}} \right)^{1/2}   
        \left( \frac{ m_3 }{4\times10^{6} \, M_{\odot}} \right)^{1/2}.
\end{equation}
 
\begin{figure}
\centering
\includegraphics[width=3in]{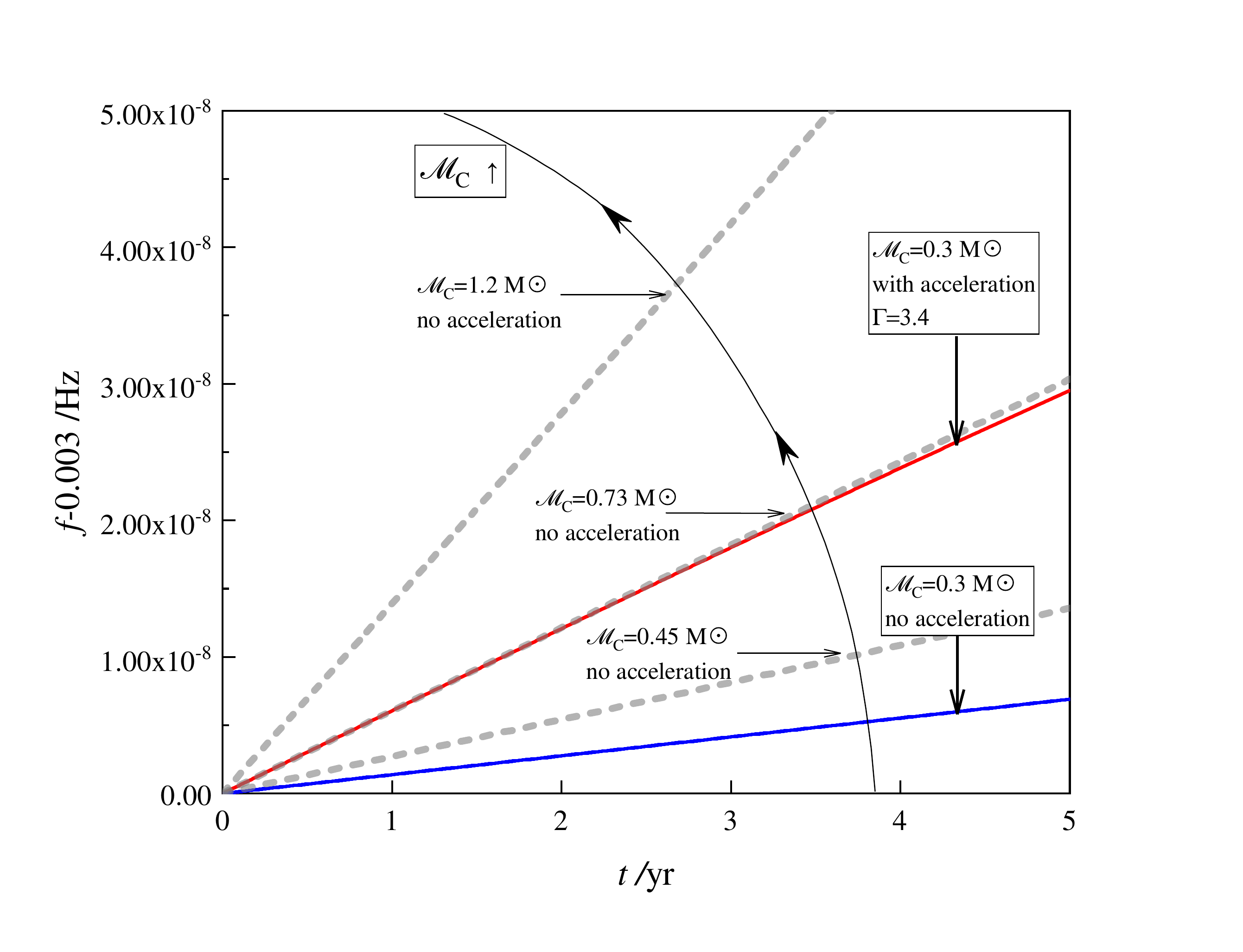}
\caption{A curved chirp signal induced by the long-term effect of peculiar
acceleration (red solid curve). It is distinguishable  from the signals of
those DWDs without acceleration (grey dashed lines).  The blue solid line is
the intrinsic chirp signal in the rest frame of the DWD.  In this example, the
tertiary has a mass of $m_3=1\,M_\odot$ and is at a distance of $r=20$ AU from
the DWD. The corresponding orbital period $P_3$ is about $69$ years and the
$\Gamma$ factor is about $3.4$.  The orbital phase $\phi$ is chosen such that
initially the acceleration effect maximizes.  }
\label{fig:curve}
\end{figure}

Second, so far we have assumed a more-or-less constant phase $\phi$ for the
outer orbit.  Such an approximation
is valid when the observational period $\tau_{\rm obs}$ is much shorter than
the orbital period of the tertiary $P_3$.  In this case, we have shown that the
peculiar acceleration contributes a constant term to $\dot{f}_o$ 
(the aforementioned $\dot{f}_{\rm acc}$).  However, if the observational period is long,
$\phi$ will sufficiently change so that $\dot{f}_o$ can no longer be regarded
as a constant. In this case, the chirp signal is not a straight line in the
$t-f_o$ diagram but curved.  Figure~\ref{fig:curve} shows this effect.
Detecting the curvature would allow us to distinguish accelerating DWDs
from those not affected by peculiar acceleration.

Suppose the peculiar acceleration induces a second time derivative 
$\ddot{f}_o$. By definition, it is the curvature of the chirp signal. 
Detecting the curvature requires that 
$\ddot{f}_{\rm acc} \tau_{\rm obs} > \Delta \dot{f}$. From this relation and
noticing that $\ddot{f}_{\rm acc} \tau_{\rm obs}\approx \dot{f}_{\rm acc} \tau_{\rm
obs}/P_3$, we derive the criterion
\begin{equation} 
r\la26 \, \rm{AU}\, \left( \frac{\rho}{100} \right)^{2/7} 
	\left( \frac{\tau_{\rm obs}}{4 \, \rm{yr}} \right)^{6/7} 
	\left( \frac{f_e}{3\, \rm{mHz}} \right)^{2/7} 
	\left( \frac{ m_3 }{1 \, M_{\odot}} \right)^{3/7} .
\label{eq: lower_limit_r}
\end{equation}
This criterion can also be derived using Equations (32), (A9) and (A10) in
\citet{2018PhRvD..98f4012R}. 
There is another way of understanding Equation~(\ref{eq: lower_limit_r}). 
Given $f_e$, $r$, $m_3$ and $\rho$, solving for $\tau_{\rm obs}$ would 
yield a minimal observational time, which we call $T_{\rm cri}$.
This is the shortest observational time that is needed to detect the
effect of peculiar acceleration.

For the DWDs in the
Galactic Centre, the criterion of distinguishing the peculiar acceleration 
becomes
\begin{equation} 
r\la0.044 \, \rm{pc}\, \left( \frac{\rho}{10} \right)^{2/7} 
	\left( \frac{\tau_{\rm obs}}{4 \, \rm{yr}} \right)^{6/7} 
	\left( \frac{f_e}{3\rm{mHz}} \right)^{2/7} 
	\left( \frac{ m_3 }{4\times10^{6} M_{\odot}} \right)^{3/7} .
\end{equation}
We have mentioned in the previous section that the mass bias could be as large
as $10{\cal M}$ if a DWD is within a distance of $0.1$ pc from the SMBH in the
Galactic Centre.  Now we can further predict that the peculiar acceleration is
undetectable if the distance is also greater than about $0.05$ pc.  In this
case, the DWD could be mis-identified as a DNS or BBH, or even a lower-mass-gap
object \citep[also see Table 1 of][]{chen20envi}. 

\begin{figure}
\centering
\includegraphics[width=3.5in]{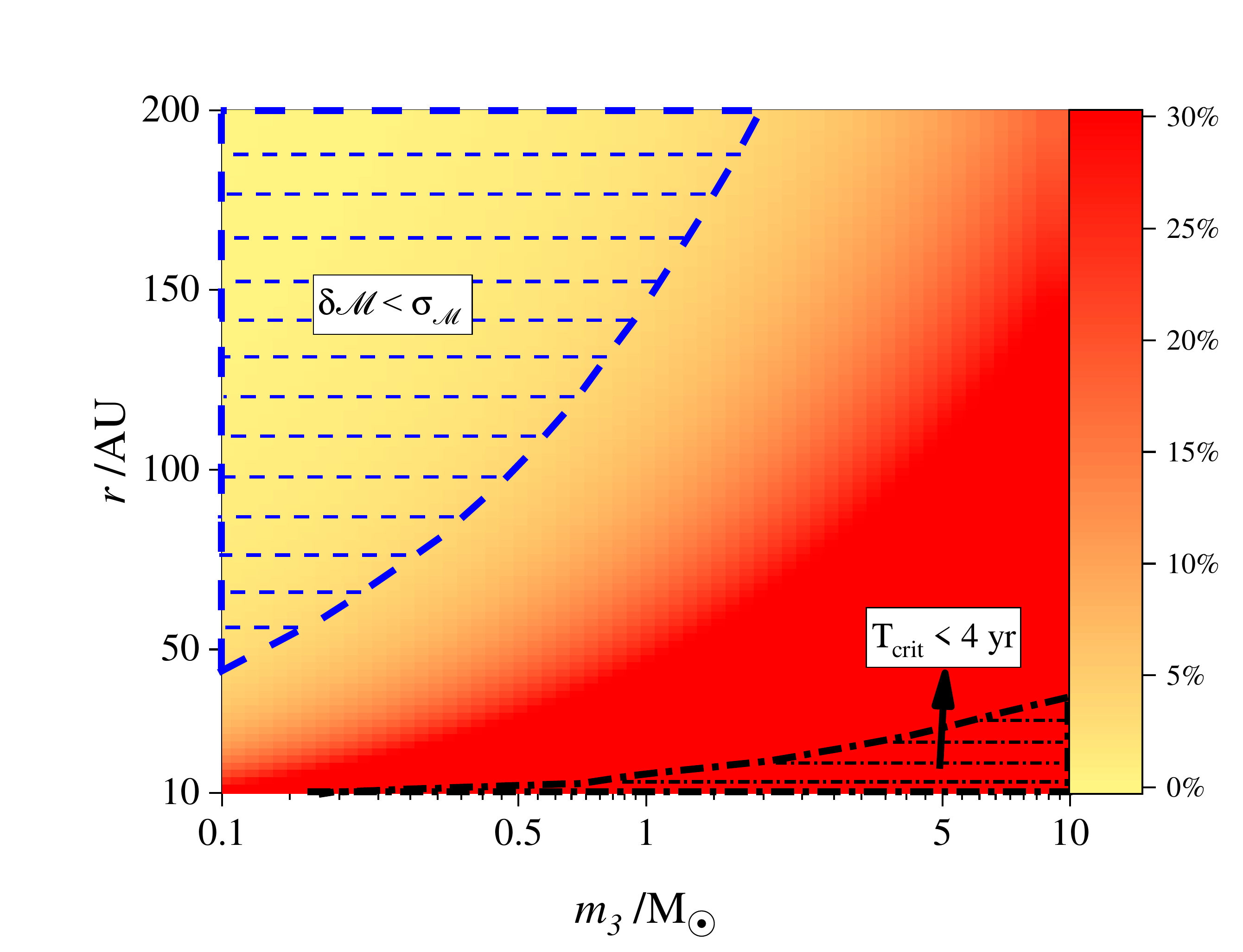}
\caption{The $m_{3}-r$ parameter space where the effect of peculiar
acceleration is unimportant (top-left hatched region labeled with $\delta {\cal
M}<\sigma_{\cal M}$) or large enough to be detectable within an observational
period of $4$ years (bottom-right hatched region labeled with $T_{\rm cri}<4$
years). The background color encodes the value of $\delta
\mathcal{M}/\mathcal{M}$. We have assumed a canonical observational scenario
where $f_e=3$ mHz, $\tau_{\rm obs}=4$ years and $\rho=100$.  }
\label{fig:range1} 
\end{figure}

Figure~\ref{fig:range1} shows the regions in the parameter space of $m_3$ and $r$
where the effect of peculiar acceleration is unimportant ($\delta {\cal
M}<\sigma_{\cal M}$) or large enough to be detectable within an observational
period of $4$ years ($T_{\rm cri}<4$ years). 

In general, wider triple systems
with lighter tertiary stars are less affected by peculiar acceleration, and
more compact triples with heavier tertiaries are more likely to reveal the
signatures of peculiar acceleration. Outside the hatched region, the effect of
peculiar acceleration is strong but difficult to discern. Here the DWDs are
likely to cause confusion. The measurement of their masses and distances could
be significantly biased.

\begin{figure}
\centering
\includegraphics[width=3.5in]{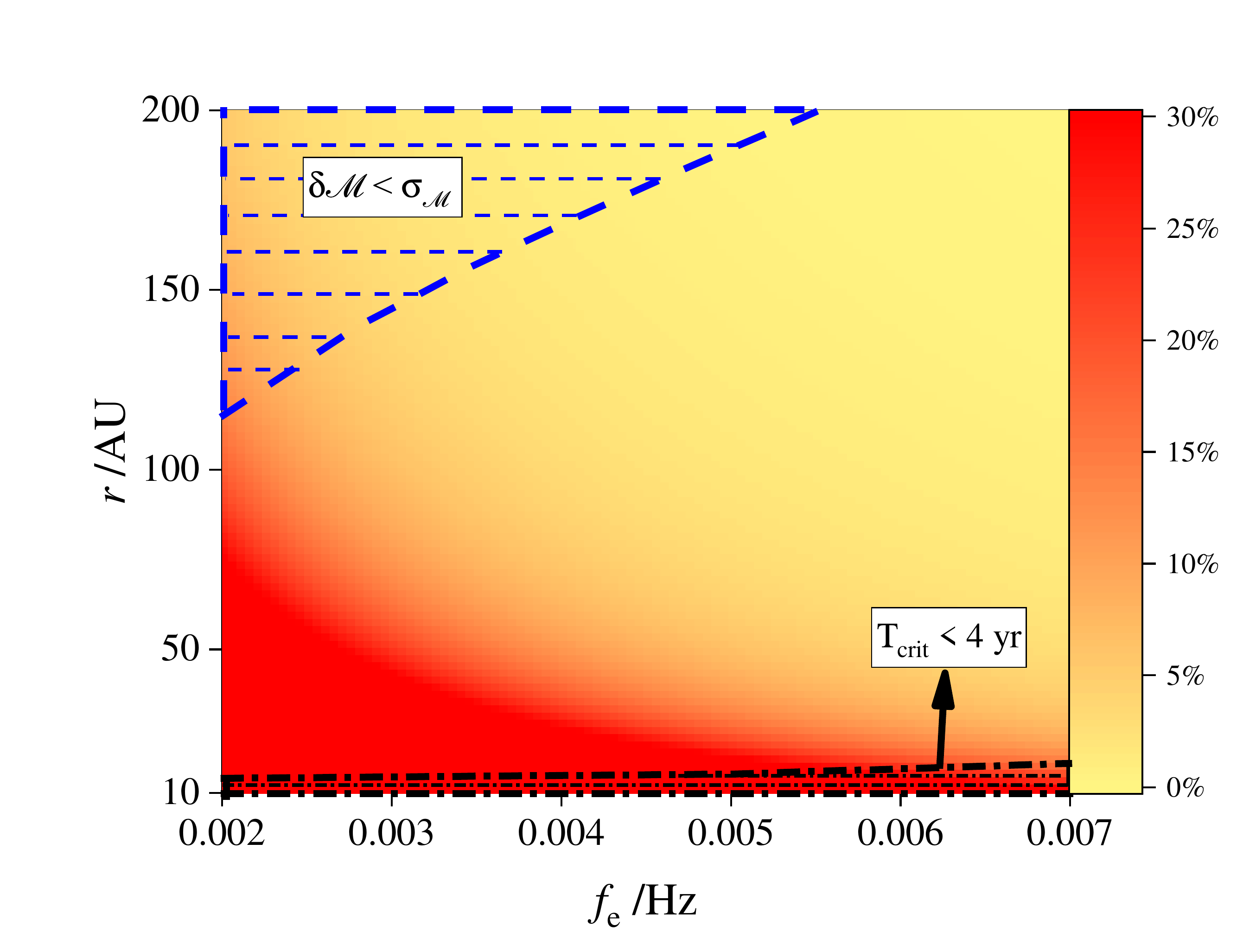}
\caption{The same as Figure~\ref{fig:range1} but showing the results in the
parameter space of $f_{e}$ and $r$. The mass of the tertiary is 
$m_{3}=1M_{\sun}$.}
\label{fig:range2}
\end{figure}

Figure~\ref{fig:range2} shows the location of the confusing DWDs in the
parameter space of $f_{e}$ and $r$. Note that the confusing DWDs reside outside
the hatched regions. In the plot we have assumed that $m_3=1\,M_\odot$. In this
case, the critical distance where confusion starts to occur is insensitive to
the GW frequency. Therefore, we find a large area above $r\ga10$ AU where the
measured ${\cal M}_o$ and $d_o$ would be biased.

\begin{figure} \centering \includegraphics[width=3.5in]{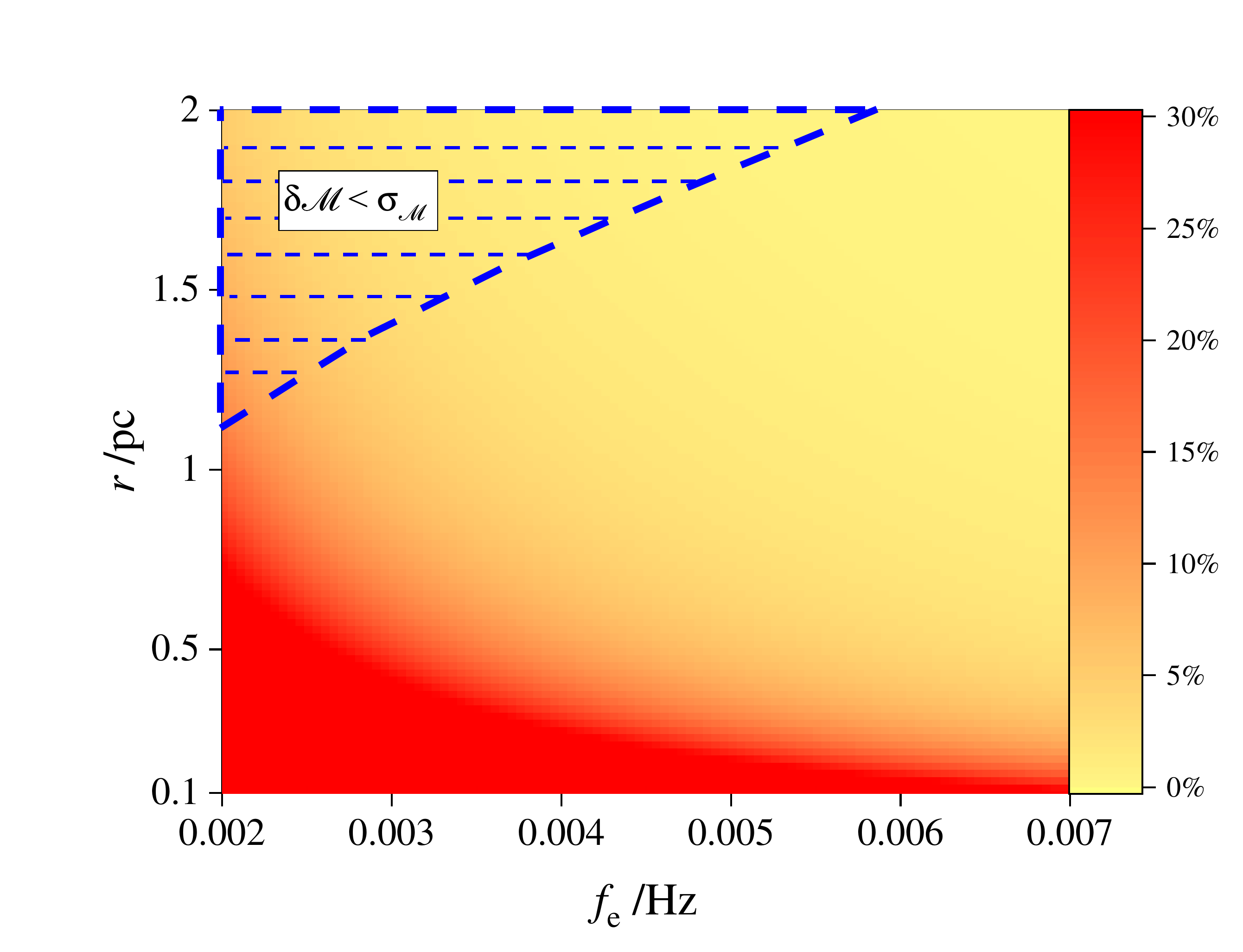}
	\caption{Similar to Figure~\ref{fig:range2} but assuming a tertiary of
	a mass of $m_3=4\times10^6\,M_\odot$. The region where $T_{\rm cri}<4$
	years is at a radius smaller than $r=0.1$ pc, and hence not shown in
the plot.} \label{fig:range3} \end{figure}

For the DWDs around the SMBH in the Galactic Centre, the result is shown in
Figure~\ref{fig:range3}. The region where the peculiar acceleration is
detectable ($T_{\rm cri}<4$) disappears. This is because in the radial range of
our interest, $r>0.1$ pc, the outer orbital period is much longer than the
observational period. In this case, the curvature of the chirp signal is
undetectable during the four years of LISA mission. Therefore, the mass is
completely degenerate with the peculiar acceleration for these DWDs in the
Galactic Centre.
 
\section{Numerical verification}
\label{sec:numerical}

Now we verify the analytical results presented in the previous two sections
using a numerical technique called ``matched filtering''
\citep{1992PhRvD..46.5236F,1994PhRvD..49.2658C}. It is commonly employed in
GW data analysis to estimate the physical parameters of the source. 
For our purpose, we use this technique to evaluate the similarity between
the waveform of an accelerating DWD and the waveform of a non-moving DWD but with a
different chirp mass. To differentiate the two waveforms, we denote the previous one
as $h_1$ and the latter one as $h_2$. 

The similarity between $h_1$ and $h_2$ can be quantified by a ``fitting
factor'' ($\rm FF$), defined as
\begin{equation}
\mathrm{FF}\equiv \frac{\left\langle h_{1} | h_{2}\right\rangle}{\sqrt{\left\langle h_{1} | h_{1}\right\rangle\left\langle h_{2} | h_{2}\right\rangle}},
\end{equation}
where $\left\langle h_{1} | h_{2}\right\rangle$ denotes the inner product
of the two waveforms, which can be computed with
\begin{equation}
	\left\langle h_{1} | h_{2}\right\rangle=2 \int_{0}^{\infty} \frac{\tilde{h}_{1}(f) \tilde{h}_{2}^{*}(f)+\tilde{h}_{1}^{*}(f) \tilde{h}_{2}(f)}{S_{\rm{n}}(f)} {\rm{d}} f.\label{eq:SNR}
\end{equation}
In the last equation, the tilde symbol means a Fourier transformation of the
waveform, the star stands for the complex conjugate and $S_{n}( f )$ is the
spectral noise density of LISA 
\citep[N2A5 configuration including a DWD background,][]{2016PhRvD..93b4003K}.  A
perfect match between the two waveforms would yield ${\rm FF}=1$.
 
As is shown in \cite{chen19gas,chen20gas}, the calculation of ${\rm FF}$ can be
performed in the time domain when the chirp rate $\dot{f}_o$ is relatively
small.  When $\dot{f}_o\tau_{\rm obs}\ll f_o$, which is the case for our DWDs, we can
use the Parseval's theorem and write
\begin{equation} \left\langle h_{1} | h_{2}\right\rangle\approx
\frac{4}{S_{n}(f)}\int_{0}^{\infty} h_{1}(t) h_{2}(t){\rm{d}} t.
\end{equation}
Consequently,
\begin{equation} \mathrm{FF}=\frac{\int_{0}^{\infty} h_{1}(t) h_{2}(t) {\rm{d}}
t}{\sqrt{\int_{0}^{\infty} h_{1}(t) h_{1}(t){\rm{d}} t \int_{0}^{\infty}
h_{2}(t) h_{2}(t) {\rm{d}} t}}.  
\end{equation}
Computing the inner product in the time domain significantly simplifies and
accelerates our calculation. 

Having defined the FF, we can use it to determine the bias $\delta {\cal M}$ induced by
peculiar acceleration. The steps are as follows. (1) Given the chirp mass ${\cal
M}$ and GW frequency $f_e$ of a DWD, we compute the waveform in the rest frame
of the binary using a 3.5 post-Newtonian approximation presented in
\citep{2009LRR....12....2S}. Although this approximation is derived for test
particles, it is appropriate for our DWDs because in the mHz band the tidal
interaction between the white dwarfs (WDs) are negligible
\citep{2017ApJ...846...95K}.  (2) We specify the mass and distance of the
tertiary, and add the effected of the peculiar acceleration to the waveform
according to Equation~(\ref{eq:fodotacc}). Such a modified waveform, as is
mentioned before, is our $h_1$. (3) We generate a suite of waveforms for
non-moving DWDs with different chirp masses ${\cal M}'$. These waveform
templates are our $h_2$. (4) We search for the maximum FF between $h_1$ and
$h_2$. The ${\cal M}'$ corresponds to the best fit is the apparent chirp mass,
${\cal M}_o$. (5) The bias is then the difference between the intrinsic ${\cal
M}$ and the observed ${\cal M}_o$.

\begin{figure*}
\begin{minipage}[t]{0.45\linewidth}
\centering
\includegraphics[width=3.3in]{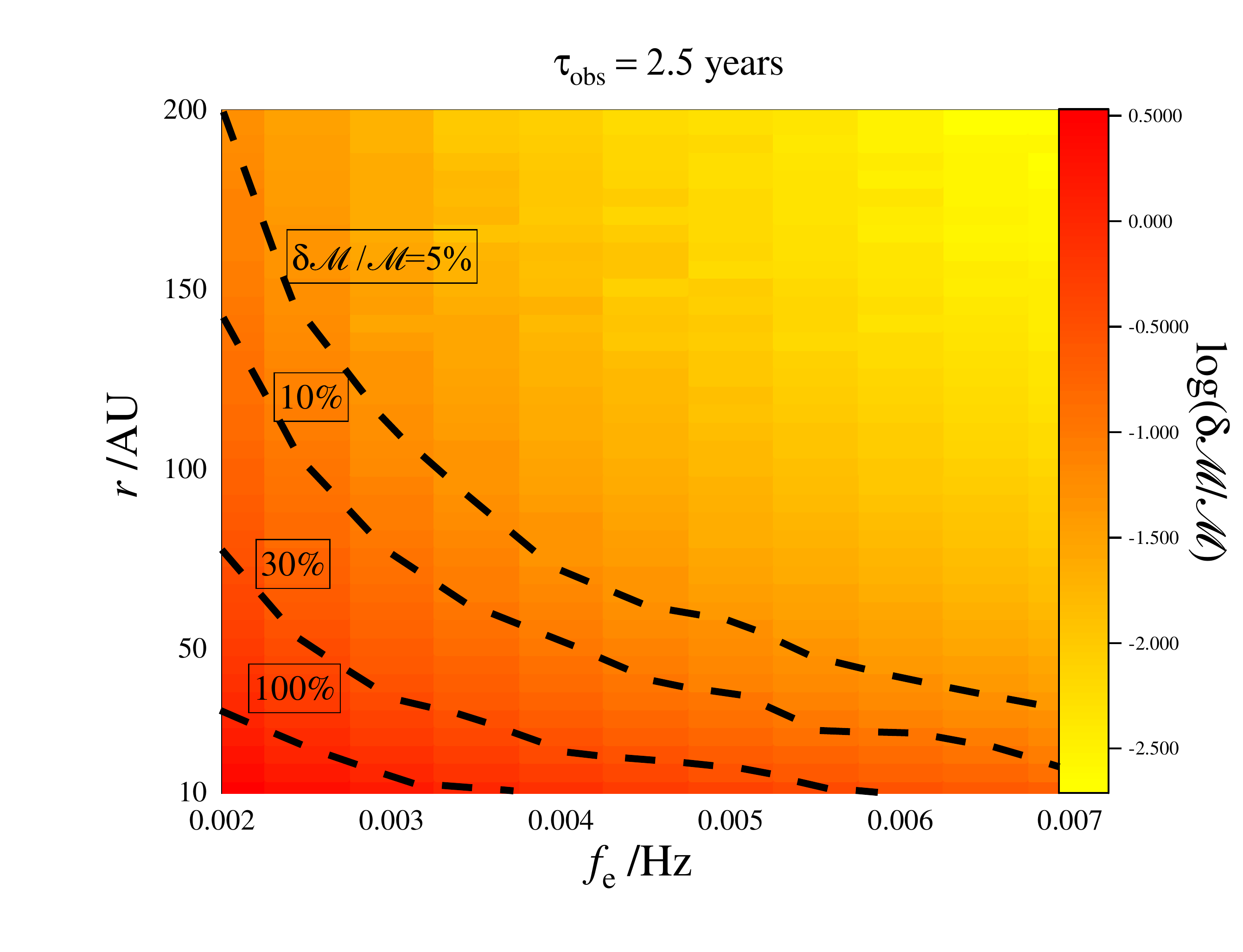}
\end{minipage}
\begin{minipage}[t]{0.45\linewidth}
\centering
\includegraphics[width=3.3in]{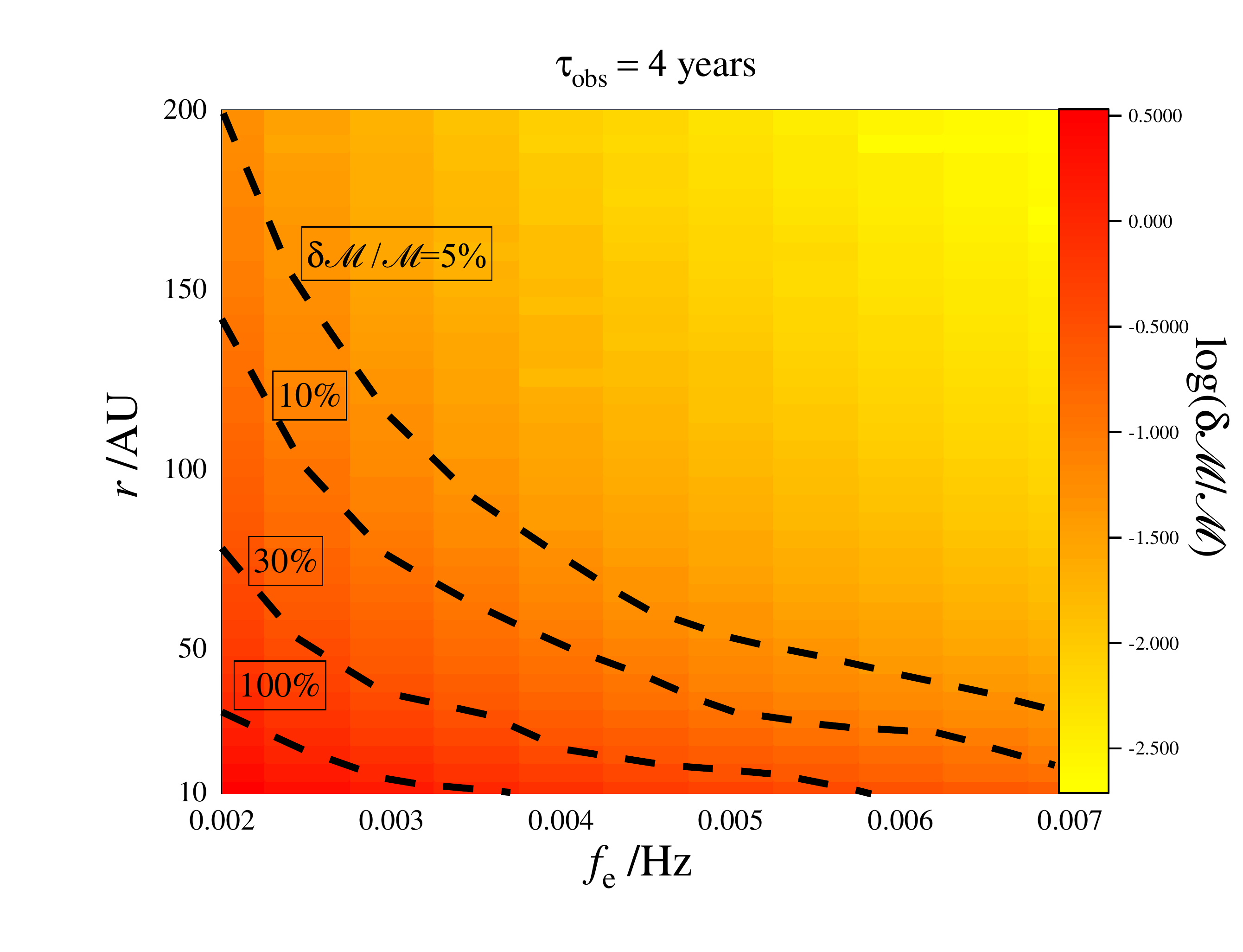}
\end{minipage}
\caption{The mass bias found by the matched-filtering method. In the left panel, the length
	of the waveforms is $2.5$ years, and in the right one the length is $4$ years.
	The other parameters are the same as in the left panel of 
	Figure~\ref{fig:analyticresult}.
}
\label{fig:contrast}
\end{figure*}

Figure~\ref{fig:contrast} shows an example of the bias found by the
matched-filtering scheme. Comparing it with the analytical result shown in the
left panel of Figure~\ref{fig:analyticresult}, we find no apparent difference.
By comparing the left and right panels in Figure~\ref{fig:contrast}, we also
find that the bias is insensitive to the observational period.  Therefore, we
are justified to use the analytical formulae presented in
Section~\ref{sec:theory} to calculate the mass bias $\delta{\cal M}$.

Next, we use the matched-filtering method to test our Equation~(\ref{eq:
lower_limit_r}). A DWD not satisfying this equation is expected to cause
confusion and leads to an wrong estimation of its chirp mass. 
In the context of matched filtering,
the condition of confusing two waveforms of different origin is
\begin{equation}
	\mathrm{FF}>1-1/(2\rho^{2}).
\end{equation}
This criterion was suggested by \citet{lindblom08} and has been used in our
earlier works \citet{chen19gas,chen20gas}. For the purpose of this work, $h_1$
is the waveform of an accelerating DWD and $h_2$ the waveform of a non-moving
DWD. The criterion to confuse $h_1$ and $h_2$ would be $\rm FF>0.99995$ if we
adopt our fiducial SNR $\rho=100$.

\begin{figure}
\centering
\includegraphics[width=3.3in]{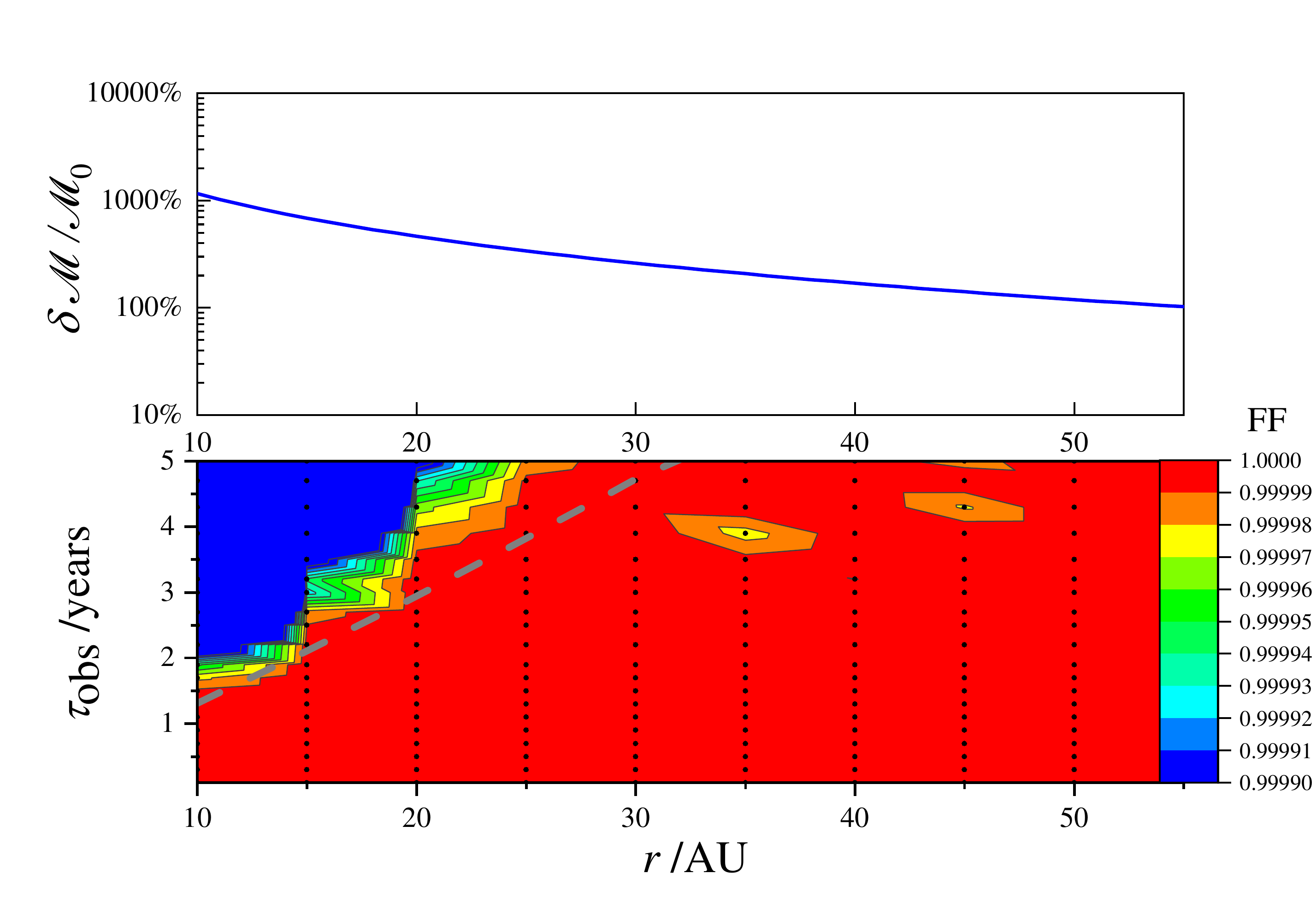}
\caption{Upper panel: The mass bias found by the matched-filtering method as a function
of the distance of the tertiary. Lower panel: Dependence of the FF on the distance of the
tertiary and the observational period. The grey dashed line is computed using 
Equation~(\ref{eq: lower_limit_r}). The black dots mark the grids where we performed
parameter estimation using the matched-filtering method.
}
\label{fig:FFresult}
\end{figure}

Figure~\ref{fig:FFresult} shows the mass bias (upper panel) and the corresponding FF derived
from the matched filtering method (lower panel).  The other models parameters are $f_e=3$
mHz, ${\cal M}=0.3\,M_\odot$, and $m_3=1\,M_\odot$. To simplify the problem, we
assume that the orbit of the tertiary is circular and the inclination is $i=0$.
Initially, the orbital phase $\phi$ is set to a value that maximizes the mass
bias. 
For comparison, our analytical criterion is shown as the grey dashed
line.  We find that the analytical and numerical criteria agree well at
$r\la15$ AU.  At $r\ga30$ AU, both results are also consistent, suggesting that
${\rm FF}\simeq1$ and LISA might confuse an accelerating DWD with a non-moving
one with a different chirp mass. The difference between the analytical and
numerical criteria is large between $r=15$ and $30$ AU. Nevertheless,  no DWD
below the dashed line has a FF lower than $0.99997$, suggesting that the
analytical criterion is more stringent than the numerical one. In the following
we will use the analytical criterion, calibrated by our numerical matched-filtering
results, to select the DWDs which might cause confusion.

\section{Population synthesis model}
\label{sec:population}

Having understood the effect of peculiar acceleration on the measurement of the
mass and distance of DWDs, we can now proceed to study  how many DWDs in the
LISA band would be substantially affected. In the following, we will focus on
the DWDs with stellar-mass companions because they are more common than those
in the Galactic Centre around the SMBH. 

We note that our current understanding of the DWD population in triples is
limited, mainly due to the observational challenges of detecting companions
within a distance of about $100$ AU from DWDs \citep{2018A&A...616A..17A,
2019MNRAS.490..657H}.  However, observations of isolated DWDs and main-sequence
multiples are relatively rich.  Therefore, we will use the information provided
by the latter observations to infer the statistical properties of the DWDs in
triples.
Given the uncertainties, it is not our desire to build a comprehensive model for
triple-star evolution. Instead, we start from the recent theoretical results
of isolated DWDs and investigate the effect of
tertiaries on the measurement of their masses and distances. 
Our model can be divided into the following three parts. 

First, we  use a Monte-Carlo code to generate a population of DWDs in the mHz
band.  The method is based on the distribution functions presented in
\citet{2019MNRAS.490.5888L}, which include the intrinsic chirp mass ${\cal M}$,
mass ratio $q:=m_2/m_1$, GW frequency $f_e$ and the real heliocentric distance
$d$ to the sun.  The initial sample contains $330,000$ DWDs. We then select
from them the binaries with a high SNR, $\rho>8$, assuming an observational
period of $4$ years by LISA.  It results in about $10,340$ DWDs in the
frequency band of $10^{-2}-10$ mHz.  The majority of these high-SNR binaries
fall in the frequency band $1-10$ mHz.  We also recover the result of
\citet{2019MNRAS.490.5888L}, that during the four-year mission of LISA about
$12,000$ DWDs can be individually resolved because of their high SNRs.
 
Second, we determine whether the $10^4$ DWDs generated in the previous step
have tertiaries. This is done using the results of the bifurcation model
presented in \cite{2009MNRAS.399.1471E}.
This model divides a stellar system
with a given total mass into smaller and smaller sub-systems, a process called
``bifurcation''.  Each step of bifurcation follows a probability which is
derived from the stellar multiples observed by the Hipparcos satellite. The
probability is a function of the bifurcation level and the mass of the system
to be divided. In general, higher bifurcation levels and smaller masses lead to
lower probability for further bifurcation. More details can be found in the
Equations (8), (17)-(20) and Table 2 in \cite{2009MNRAS.399.1471E}.

Following the bifurcation model, we generate a sample of about $2\times10^7$
binary and triple main-sequence stars. The sample size is far less than the
real number of binaries and triples in the Milky Way, but large enough for our
purpose, i.e., deriving the probability that a DWD generated in the first step
has a tertiary companion. Not all of the $2\times10^7$ stellar multiples 
can produce close DWDs in the
LISA band. Therefore, we apply two more criteria to find the systems which
could yield LISA DWDs. (1) The stellar components of the binaries (including
the inner binaries in the triples) fall in the mass range of $0.9M_{\odot}\sim
8M_{\odot}$. This criterion ensures that the main-sequence binaries eventually
evolve into DWDs. It is based on the empirical relationship 
\begin{equation} 
M_{\rm{core}} = 0.1 \left(
	M_{\rm{MS}}/M_{\odot} \right)^{1.4}\label{eq:ms} 
\end{equation}
between the mass of the progenitor star
$M_{\rm{MS}}$ and the mass of the stellar core $M_{\rm{core}}$
\citep{2014LRR....17....3P}.  (2) The orbital
periods of the main-sequence binaries (including the inner binaries) fall in the range
between $0.1$ day and $1$ year. This requirement comes from the fact that 
for a binary star with a total mass of about $2\,M_\odot$,
the
orbital semimajor axis typically shrinks by a factor of $100$ during the later
common-envelope phases
\citep{2014LRR....17....3P,2013MNRAS.430.2262H,2012ApJ...744...12W}. In this
case, an initial orbital period of $10^{-3}-1$ year would lead to a final
orbital frequency bewteen $10^{-2}$ and $10$ mHz. By applying the above two
selection criteria, we have about $10^5$ systems left. About $36\%$ are
binaries and $64\%$ are triples. This result is also consistent with the
observational result that the binaries with shorter orbital periods are more likely
in triples \citep{2006AJ....131.2986P,2008msah.conf..129T}.  Based on these
results, we randomly select $6,400$ DWDs from the previous sample of $10,340$ 
and assume that they have tertiaries. 

\begin{figure*}
\begin{minipage}[t]{0.45\linewidth}
\centering
\includegraphics[width=3in]{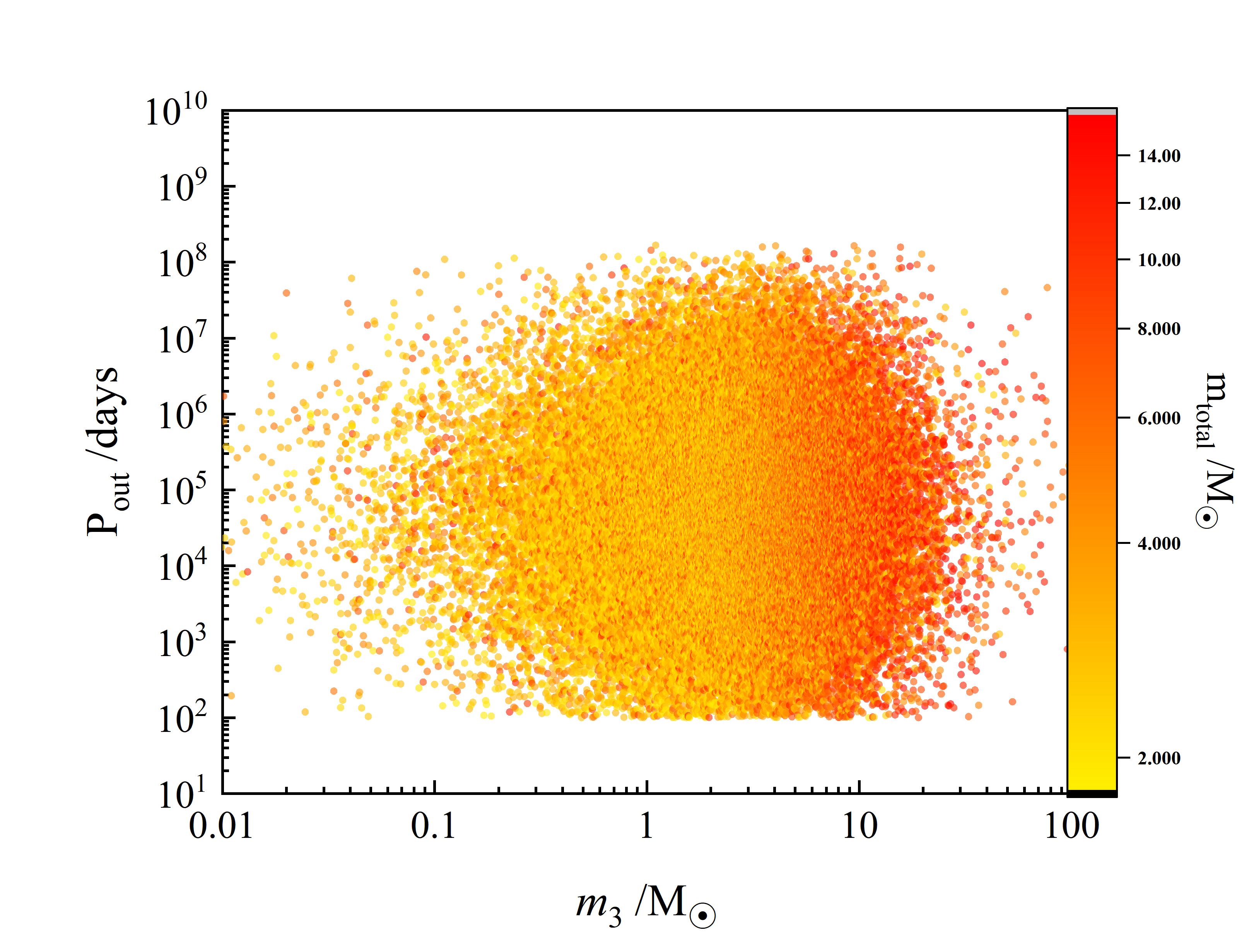}
\end{minipage}
\begin{minipage}[t]{0.45\linewidth}
\centering
\includegraphics[width=3in]{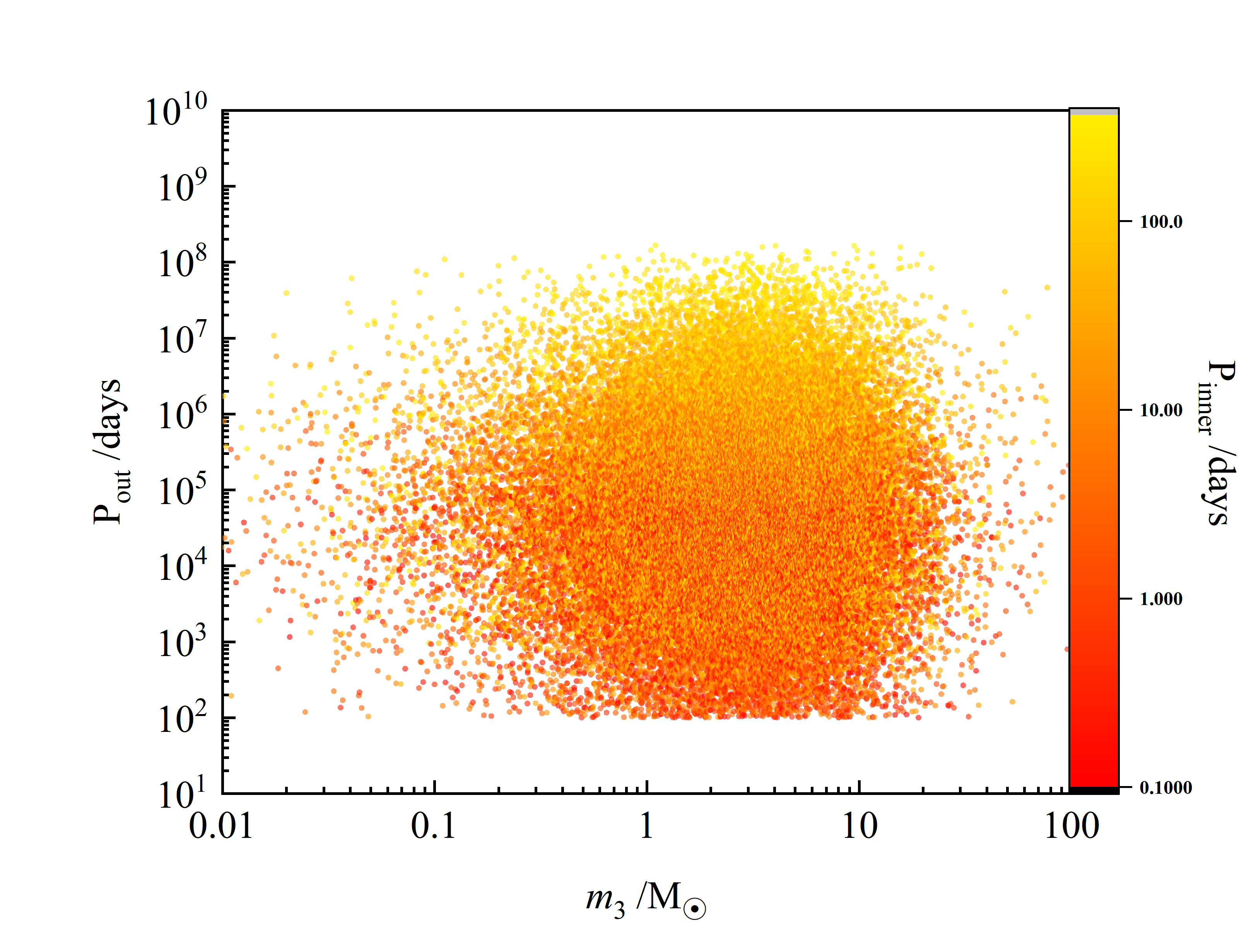}
\end{minipage}
\caption{Distribution of the $6.4\times10^4$ triple main-sequence stars which will
produce DWDs in the LISA band. Here $m_3$ is the initial mass of the tertiary star
and $P_{\rm out}$ stands for the initial orbital period of the outer binary.
In the left panel the color indicates the total mass $m_{\rm total}$ of the inner 
main-sequence binary,
and in the right panel the color stands for the initial orbital period of the inner binary.
}
\label{fig:MS population}
\end{figure*}

Third, for each of the $6,400$ selected DWDs with tertiaries, we determine the mass and
distance of the third star. This is done in two steps. (1) Given a DWD with a
tertiary, we identify the progenitor triple based on the 
$6.4\times10^4$  
triple stars from our bifurcation model. The distribution of the $6.4\times10^4$
triples is shown in
Figure~\ref{fig:MS population}. 
We find that tertiaries are more likely to appear for heavier binaries (left panel) and 
their orbital periods are closely correlated with the orbital periods of the inner
binaries (right panel). Because of these correlations, the mass and distance of a tertiary should not
be selected randomly.
In our model, the progenitor triple is identified 
according to the masses of the binary components and the relationship given by
Equation~(\ref{eq:ms}). Sometimes more than one progenitor triple can be found
for a DWD with a tertiary. In these cases, we randomly select one progenitor.
(2) Having found the progenitor triples, we evolve the outer
orbits according to the ages of the $6,400$ DWDs. The age of the DWD is randomly
selected between the age of the progenitor stars, $10\,(M_{\rm MS}/M_\odot)^{-3}$
Gyr \citep{kippenhahn94}, and the age of the Galaxy, $13.7$ Gyr. The evolution
of the outer triple is caused by the mass loss of the progenitors stars during their post-main-sequence
evolution, and we use the equation for adiabatic orbital expansion, 
\begin{equation}
\frac{r(\rm{now})}{r(\rm{initial})} = \frac{m_1+m_2+m_3+\Delta m}{m_1+m_2+m_3}
\label{eq:mass loss}
\end{equation}
\citep{2016ComAC...3....6T}, to find the current distance of the tertiary.  The
amount of mass loss, $\Delta m$, can be calculated using the relationship given
by Equation~(\ref{eq:ms}). Note that if the age of the DWD exceeds the
main-sequence lifetime of the tertiary star, the tertiary also loses mass and 
evolves into a
compact object. In this case, we add the mass loss to $\Delta m$ and update
$m_3$ according to Equation~(\ref{eq:ms}).   

\section{Mock LISA observation}
\label{sec:observation}

So far we have generated a sample of about $10^4$ DWDs in the LISA band, whose
masses, heliocentric distances and GW frequencies follow the distributions
given in \citet{2019MNRAS.490.5888L}.  The difference of our sample is that
about $64\%$ of them are in triple systems, and we have determined the
masses and distances of the tertiaries. 

Then we check whether LISA could detect the chirp signals of these $10^4$ DWDs.
We first calculate the GW amplitude in the frequency domain for each DWD
\citep[following][]{2002MNRAS.333..469S} and use Equation~(\ref{eq:SNR}) to
compute the SNR ($\rho$). A canonical observational period of $\tau_{\rm
obs}=4$ years is assumed.  For those isolated DWDs, which constitute $36\%$ of
the sample, we use Equation~(\ref{eq:resolve}) to select the ones with
detectable chirp signals.  For the rest $64\%$, which are triples, we use the
criterion $|\dot{f}_o|>\Delta\dot{f}$ to take into account the effect of
peculiar acceleration.  In the end, we have $8105$ DWDs whose chirping rate is
resolvable by LISA. We note that this number is about eight times larger than
the number of DWDs with measured mass given in \citet{2019MNRAS.490.5888L},
because they used a more strict requirement that the chirp mass should be
measured to an accuracy of $10\%$. 

For these $8,105$ DWDs, we calculate their apparent chirp masses and apparent
distances using Equations~(\ref{eq:Mo}), (\ref{eq:do}) and (\ref{eq:Gamma1}),
so that we can evaluate the biases relative to the real masses and real
distances. In the calculation, for each triple system we have assumed a random
distribution for the inclination ($i$) and phase ($\phi$) of the outer orbit,
so that the effect of peculiar acceleration is, on average, weaker than what is
predicted by Equation~(\ref{eq:Gamma1}).  We also apply Equation~(\ref{eq:
lower_limit_r}) to judge whether we would  confuse the DWD in a triple 
and an isolated DWD with a different chirp mass.

\begin{figure*} \begin{minipage}[t]{0.45\linewidth} \centering
	\includegraphics[width=3in]{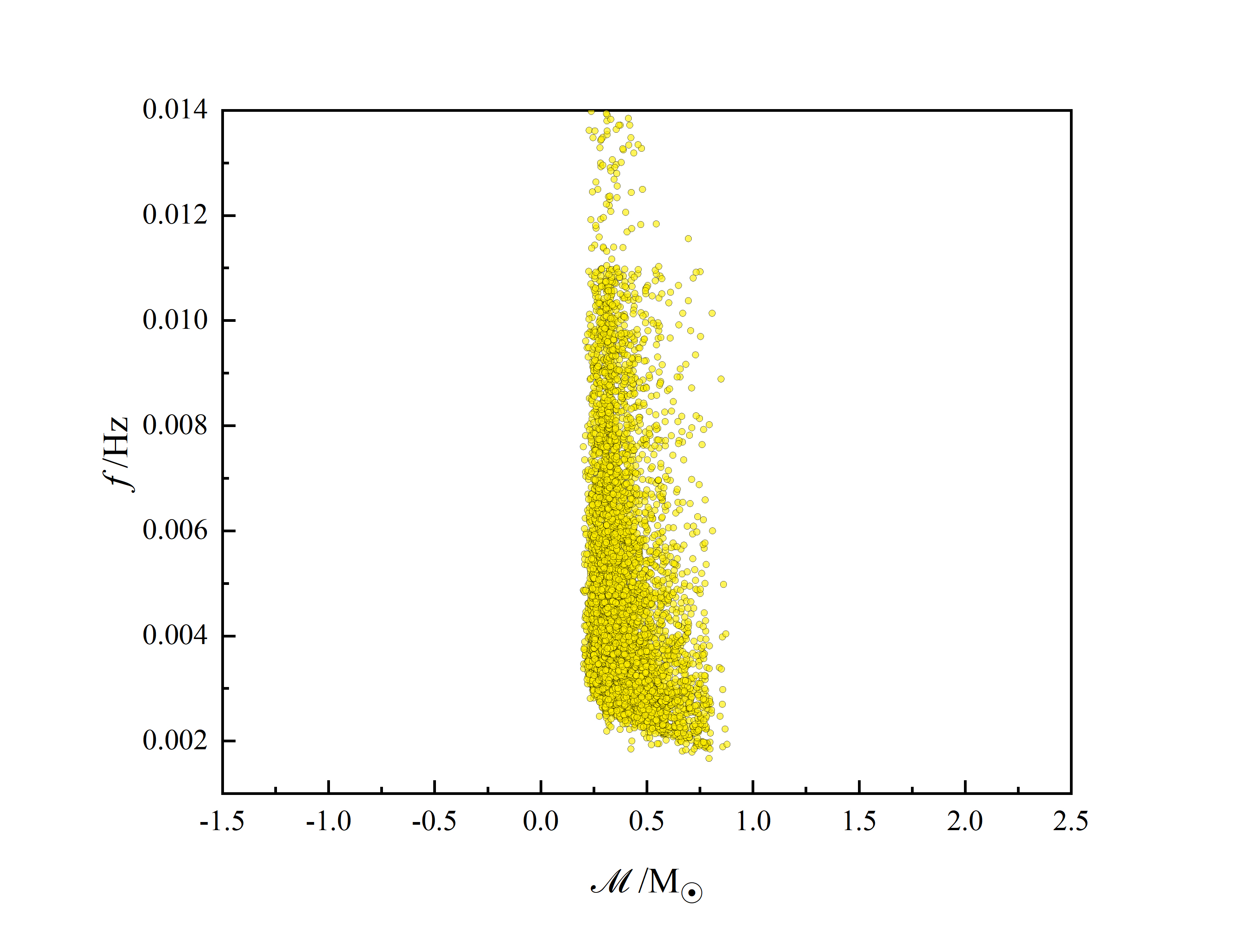} \end{minipage}
	\begin{minipage}[t]{0.45\linewidth} \centering
	\includegraphics[width=3in]{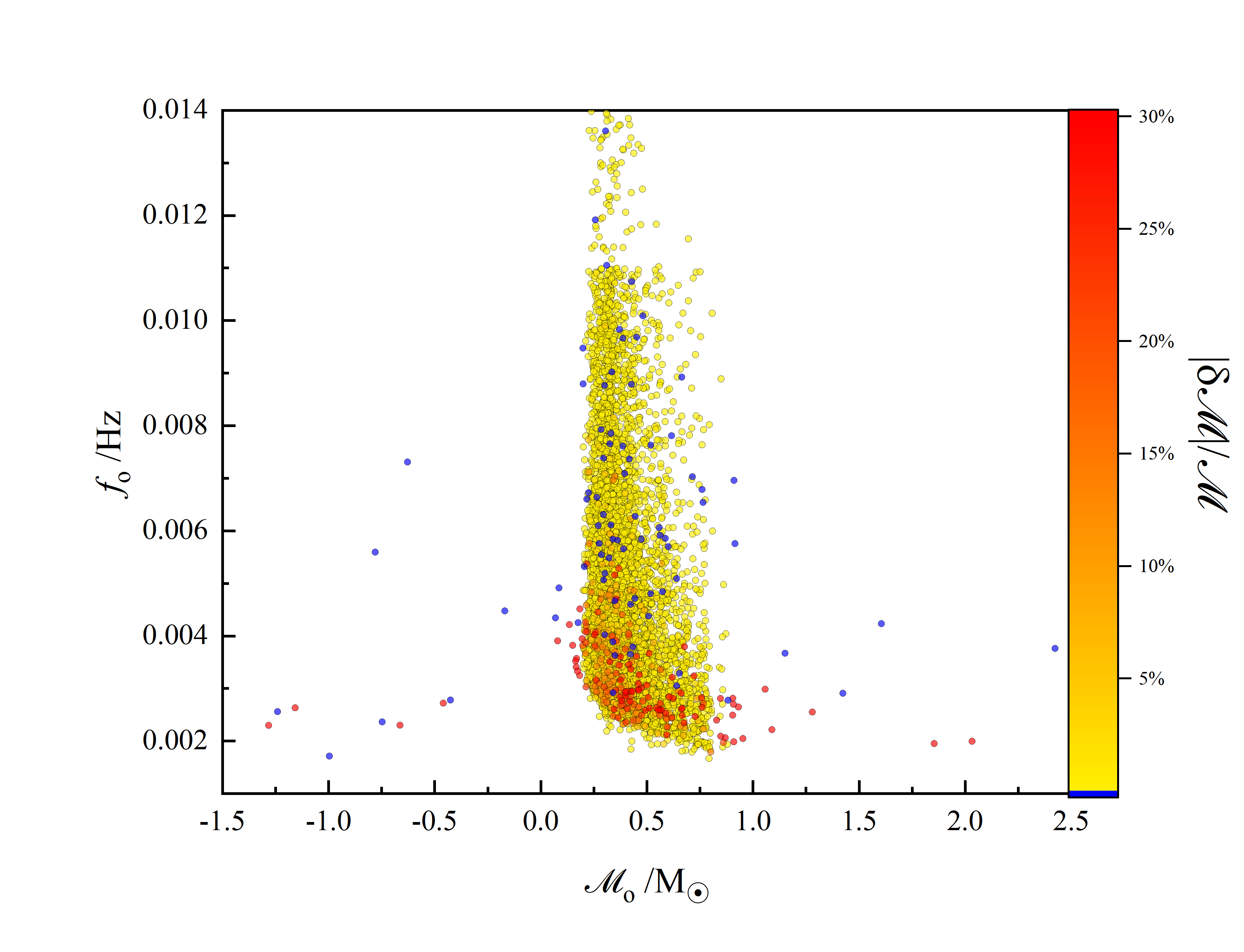} \end{minipage}
	\caption{Left panel: The intrinsic chirp masses and GW frequencies of
	the $8105$ DWDs whose chirping rate can be resolved by LISA. Right
	panel: The apparent chirp masses and GW frequencies of the same DWDs,
	color-coded by the mass bias. The blue dots are the DWDs with a
	distinguishable peculiar acceleration. Several DWDs have negative chirp
	masses because the $\Gamma$ factor
	becomes smaller than $-1$.} \label{fig:affected parameter}
	\end{figure*}

Figure~\ref{fig:affected parameter} shows the real chirp mass and frequency
(left panel) of the $8,105$ DWDs and the observed values (right panel).
Comparing the two panels, we find that the majority of the DWDs are not
significantly affected by the presence of tertiaries (yellow dots).  The
DWDs most affected by peculiar acceleration (red dots) are those with relatively low GW
frequencies, e.g., below $4$ mHz. Most of them are indistinguishable
from isolated, non-moving DWDs.  For these confusing DWDs, the intrinsic chirp
rate $\dot{f}_e$ is small, so that it can be overcome by the additional chirp
rate $\dot{f}_{\rm{acc}}$ induced by peculiar acceleration.  Moreover, we find
from the right panel that at higher frequencies there are less red dots but
more blue ones.  This trend indicates that at high frequencies, it is difficult
for a tertiary to induce a large mass bias in a DWD and at the same time keep
the accelerating DWD indistinguishable from isolated, non-moving DWDs. This is
so because higher frequency corresponds to larger $\dot{f}_{\rm{GW}}$, so that
the tertiary should get closer to the DWD to induce a significant mass bias.
As the tertiary gets closer, the period of the outer orbit shortens and the
curvature of the chirp signal becomes more prominent.
  
Several confusing DWDs (red dots) in the right panel of
Figure~\ref{fig:affected parameter} are worth mentioning.  
\begin{itemize}
\item One red dot appears at ${\cal M}_o>2.0\,M_\odot$ with a SNR of
$8.6$.  Therefore, it looks like a DNS, e.g., $m_1=m_2=2.3\,M_\odot$, or even a
binary containing a lower-mass-gap object, e.g., $m_1=4\,M_\odot$ and
$m_2=1.4\,M_\odot$.  In fact, the real chirp mass is only $0.39\,M_\odot$. The
apparent chirp rate is raised by a tertiary stellar-mass black hole of a mass
of $7.4\,M_\odot$ orbiting at a distance of $22$ AU from the DWD.  The real
heliocentric distance of this system is about $8.4$ kpc, but due to the
peculiar acceleration, it appears to be at a distance of $130$ kpc. 

\item Less extreme cases can be found at ${\cal M}_o\simeq1\,M_\odot$. For
example, one red dot appears at ${\cal M}_o\simeq1.28\,M_\odot$ with a SNR of
$12$, so that it looks like a DNS.  The real chirp mass is actually
$0.35\,M_\odot$. The tertiary is a compact object at a distance of $14$ AU and has a mass of
$1.4\,M_\odot$.  Therefore, it is either a high-mass WD or a low-mass neutron
star. The real heliocentric distance is about $14$ kpc, but it appears to be at
$d_o\simeq126$ kpc because of the peculiar acceleration.

\item There are $11$ red dots with a chirp mass significantly smaller than
$0.2\,M_\odot$. Theoretically, such small compact-object binaries are
difficult to form. In fact,  their
real chirp masses are about $0.3\,M_\odot$. The tertiaries
include main-sequence stars, neutron stars and black holes.
The most extreme one has an apparent chirp mass of
${\cal M}_o\simeq0.079\,M_\odot$ with a SNR of $46$. 
Such a system could be
misidentified as a binary of primordial black holes. Its real chirp mass is 
$0.29\,M_\odot$, and the tertiary is a black hole of $3.5\,M_\odot$ orbiting at a
distance of $35$ AU. The real heliocentric distance is $5.6$ kpc, but because of the
peculiar acceleration, it appears closer, at about $d_o\simeq650$ pc.  
 
\item We also find red dots with negative chirp masses. They are caused by
the fact that $\dot{f}_{\rm acc}$, and hence the total $\dot{f}_o$, can both be
negative. The signal with a negative chirp rate is known as ``inverse chirp''.
In the conventional model of DWD, such a signal can be produced by a mass
transfer between the two WDs \citep{2004MNRAS.349..181N}. 
Because of this possibility, 
it is difficult to distinguish these red dots with  ${\cal M}_o<0$ from the DWDs undergoing mass transfer.
We note that most of these confusing DWDs have black-hole or neutron-star
tertiaries. 
\end{itemize}

\begin{figure}
\centering
\includegraphics[width=3in]{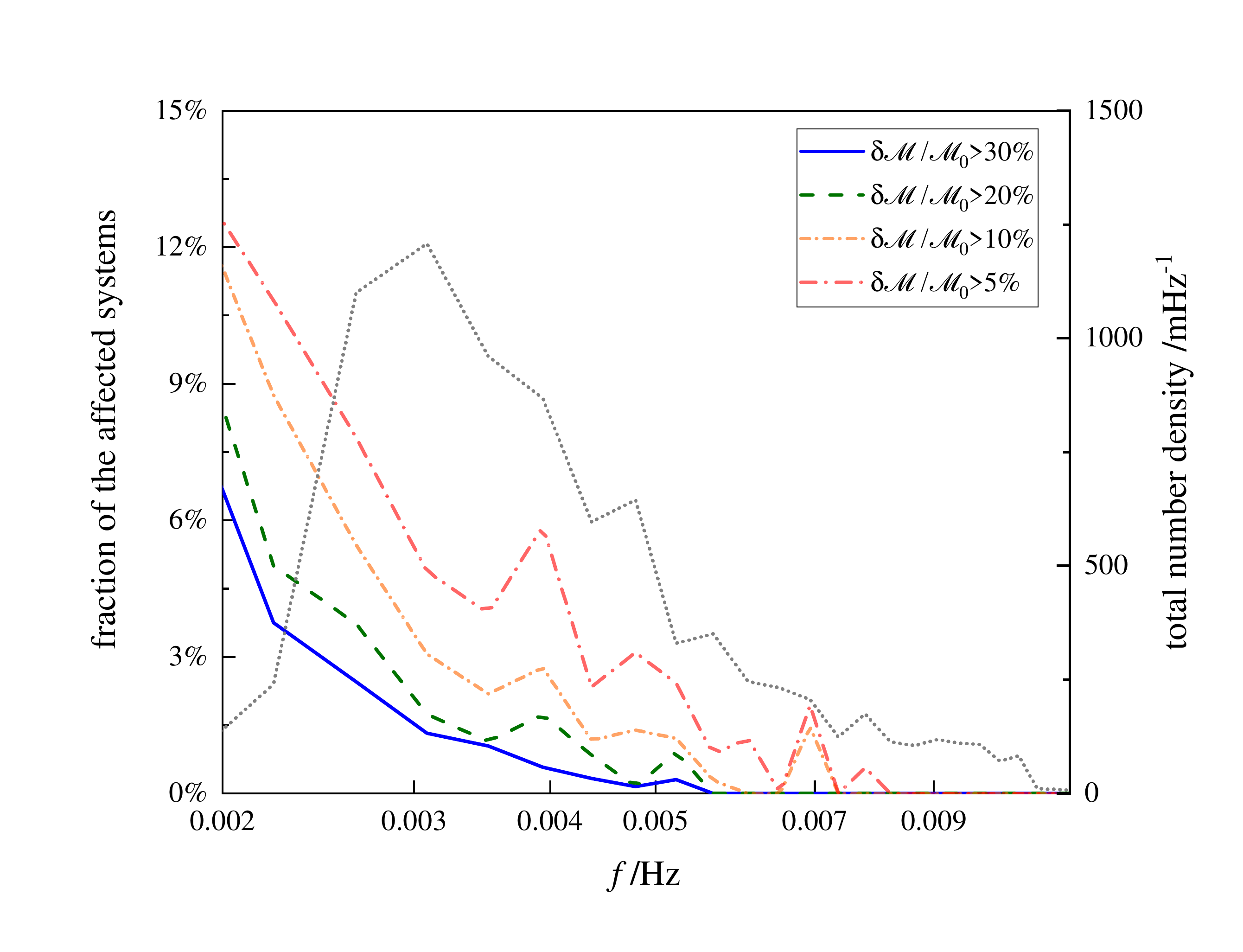}
\caption{The fraction of confusing DWDs in each frequency bin (colored curves).
The red, orange, green and blue lines, respectively, refer to the DWDs with a mass bias
greater than $5$, $10$, $20$ and $30$ per cent. For reference, the black dotted line shows the
distribution of all the DWDs with measurable chirp masses.
}
\label{fig:affected fraction}
\end{figure}

Among these $8,105$ DWDs of which we can derive chirp masses,
$75,\,144,\,193,\,319$ have a mass bias greater than $30,\,20,\,10,\,5$ per
cent and at the same time they are indistinguishable from isolated DWDs.
Therefore, confusing DWDs  constitute about $9\%$ of the LISA DWDs with
measurable masses.  Figure~\ref{fig:affected fraction} shows more clearly the
distribution of these confusing DWDs in the frequency space. We find that in
general there is a larger fraction of confusing DWDs at lower frequencies.
This result can be understood since the chirp rate due to GW radiation drops as
the frequency decreases. 

\begin{figure}
\centering
\includegraphics[width=3in]{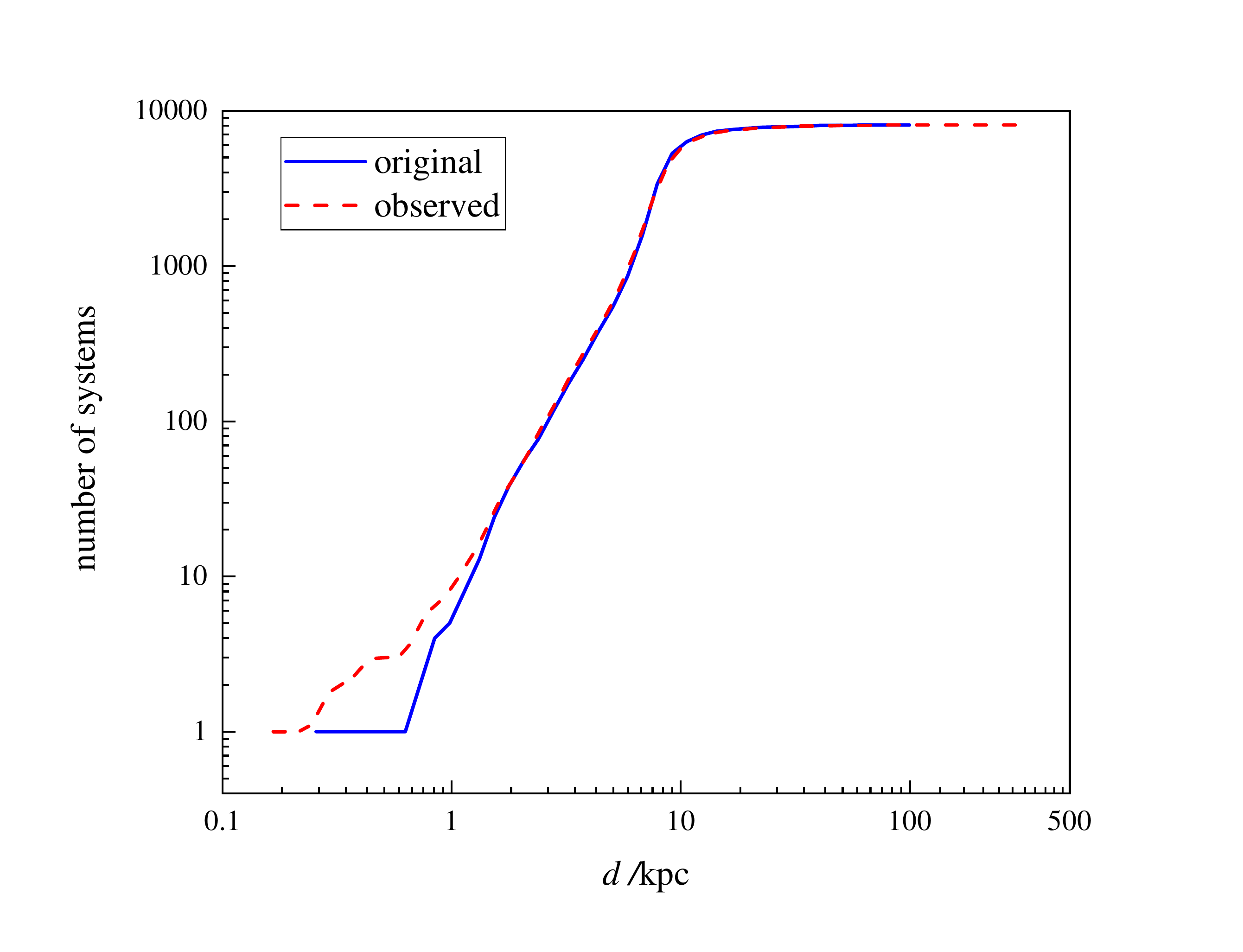}
\caption{Cumulative distribution of the real (blue solid line) and observed (red dashed line) 
distances for the $8\times10^3$ DWDs with measurable chirp masses.}
\label{fig:distance}
\end{figure}

We have shown in Section~\ref{sec:theory} that bias in the measurement of
chirp mass will also result in a bias in the measured distance.
Figure~\ref{fig:distance} shows this effect using the $8,105$ simulated DWDs.
Comparing the red and blue curves, we find an significant excess of binaries at
a relatively low distance, below $1$ kpc, when triples and the resulting
peculiar acceleration are taken into account.  Since these additional binaries
are fake and their real distances are much larger than the apparent ones, it
would be impossible for future deep photometric surveys, such as LSST and Gaia
\citep[e.g.][]{2019MNRAS.483.5518K}, to find their electro-magnetic
counterparts.  Moreover, Figure~\ref{fig:distance} also shows that there are
binaries in the LISA band which appear to be beyond $100$ kpc. They are also
fake since the real distance (blue curve) stops at $100$ kpc.  These results
highlight the potential problem that the DWDs in triples may bring to our
understanding of the Galactic structure.

\section{Summary and conclusion}
\label{sec:conclusion}

In this work, we have demonstrated the effect of peculiar acceleration induced
by tertiaries on the measurement of the mass and distance of the DWDs in the
LISA band. Different from the previous works, our method is mainly analytical,
but we also verify the key results using numerical simulations. Using the
analytical results, we are able to conduct mock LISA observation of a large
population of simulated DWDs with tertiaries.

We find that about $9\%$ of the DWDs of which LISA can measure a chirp mass are
affected by peculiar acceleration.  They reside mostly in the frequency band of
$2-4$ mHz and are indistinguishable from isolated, non-moving DWDs.  The bias
in the mass measurement is mostly $(5-30)\%$, but there are also cases in which
the bias exceeds $100\%$. In those extreme cases, the DWDs can mimic DNSs,
BBHs, DWDs undergoing mass transfer, or even binaries containing 
lower-mass-gap objects and primordial black
holes. As a result of the bias, the distance of the DWDs measured by LISA are
also inaccurate. It leads to an apparent over-population of DWDs within a
heliocentric distance of $1$ kpc as well as beyond $100$ kpc.

In general, our results highlight the importance of modeling the astrophysical
environments of GW sources \citep[e.g.][]{chen20envi}. For DWDs, 
in particular, developing population synthesis models for
triple stars is necessary and urgently needed.

\section*{Acknowledgements}

This work is supported by NSFC grants No. 11721303 and 11873022. X.C.  is
supported partly by the Strategic Priority Research Program “Multi-wavelength
gravitational wave universe” of the Chinese Academy of Sciences (No.
XDB23040100 and XDB23010200). The computation in this work was performed on the
High Performance Computing Platform of the Centre for Life Science, Peking
University.






\bibliographystyle{mnras}
\bibliography{bibbase,mybib}




\bsp	
\label{lastpage}
\end{document}